\documentclass{aa}  

\usepackage{graphicx}
\usepackage{txfonts}
\usepackage{amsmath}
\usepackage[dvipsnames]{xcolor}
\usepackage[]{hyperref}   
\hypersetup{colorlinks=true,urlcolor=MidnightBlue,hypertexnames=true,plainpages=false,naturalnames=true,pdftitle={28 Years of Sun-as-a-star Ultraviolet Light Curves from SOHO},pdfauthor={E. Sandford},linkcolor=WildStrawberry,citecolor=ForestGreen}
\usepackage{orcidlink}
\usepackage{makecell}

\makeatletter
\renewcommand*\aa@pageof{, page \thepage{} of \pageref*{LastPage}}
\makeatother

\begin{document}

   \title{28 Years of Sun-as-a-star Extreme Ultraviolet Light Curves from SOHO EIT}

   \author{Emily Sandford
          \inst{1}\orcidlink{0000-0003-0822-1839},
          Fr\'{e}d\'{e}ric Auch\`{e}re
          \inst{2}\orcidlink{0000-0003-0972-7022}
          \and
          Annelies Mortier
          \inst{3}\orcidlink{0000-0001-7254-4363}
          \and
          Laura A. Hayes
          \inst{4}\orcidlink{0000-0002-6835-2390}
          \and
          Daniel M\"{u}ller
          \inst{5}\orcidlink{0000-0001-9027-9954}
          }

   \institute{Leiden Observatory, University of Leiden, Einsteinweg 55, NL-2333 CC Leiden, The Netherlands\\
         \email{sandford@strw.leidenuniv.nl}
         \and
         Universit\'{e} Paris-Saclay, CNRS, Institut d’Astrophysique Spatiale, 91405, Orsay, France
         \and
         School of Physics \& Astronomy, University of Birmingham, Edgbaston, Birmingham B15 2TT, UK
         \and
         Astronomy $\&$ Astrophysics Section, School of Cosmic Physics, Dublin Institute for Advanced Studies, DIAS Dunsink Observatory, Dublin D15XR2R, Ireland
         \and
         European Space Agency, ESTEC, P.O. Box 299, NL-2200 AG Noordwijk, The Netherlands.\\
             }

   \date{Received September 15, 202X; accepted September 15, 202X}

 
  \abstract
   {The Solar and Heliospheric Observatory (SOHO) Extreme-ultraviolet Imaging Telescope (EIT) has been taking images of the Solar disk and corona in four narrow EUV bandpasses (171\AA, 195\AA, 284\AA, and 304\AA) at a minimum cadence of once per day since early 1996. The time series of fully-calibrated EIT images now spans approximately 28 years, from early 1996 to early 2024, covering solar cycles 23 and 24 in their entirety, as well as the beginning of cycle 25.}
   {We aim to convert this extensive EIT image archive into a set of `Sun-as-a-star' light curves in EIT's four bandpasses, observing the Sun as if it were a distant point source, viewed from a fixed perspective. }
   {To construct the light curves, we sum up the flux in each EIT image into one flux value, with an uncertainty accounting for both the background noise in the image and the potential spillover of flux beyond the bounds of the image (which is especially important for the bands with significant coronal emission). We correct for long-term instrumental systematic trends in the light curves by comparing our 304\AA\ light curve to the ultraviolet light curve taken by SOHO's CELIAS/SEM solar wind monitoring experiment, which has a very similar bandpass to the EIT 304\AA\ channel. We correct for SOHO's viewing angle by fitting a trend to the flux values with respect to SOHO's heliocentric latitude at the time of each observation.}
   {We produce two sets of Sun-as-a-star light curves with different uncertainty characteristics, available for download at \href{https://doi.org/10.5281/zenodo.15474179}{https://doi.org/10.5281/zenodo.15474179}, either of which might be preferred for different types of future analysis. In version (1), we treat the EIT instrumental systematics consistently across the entire SOHO mission lifetime, producing a light curve with approximately homoscedastic uncertainties. In version (2), we only divide out the EIT instrumental systematics from 12 November 2008 onward; this is the point at which these systematics start to have a noticeable deleterious effect on the data. Version (2) therefore has heteroscedastic uncertainties, but these uncertainties are much smaller than the version (1) uncertainties over the first half of the mission.}
   {We find that our EUV light curves trace the Sun's $\sim 11$-year solar activity cycle and $~\sim 27$-day rotation period much better than comparable optical observations. In particular, we can accurately recover the solar rotation period from our 284\AA\ light curve for 26 out of 28 calendar years of EIT observations (93\% of the time), compared to only 3 out of 29 calendar years (10\% of the time) for the VIRGO total solar irradiance time series, which is dominated by optical light. Our EIT light curves, in conjunction with Sun-as-a-star light curves at optical wavelengths, will be valuable to those interested in inferring the EUV/UV character of stars with long optical light curves but no intensive UV observations, as well as to those interested in long-term records of solar and space weather.}


   \maketitle
%

\section{Introduction}
\label{sec:intro}
The Solar and Heliospheric Observatory (SOHO; \citealt{domingo1995}) is a space-based heliophysics observatory jointly developed and operated by ESA and NASA. SOHO launched in December 1995, carrying a suite of instruments for photometric, spectroscopic, and in-situ observations of the Sun and the solar wind. Many of these instruments are still operating. SOHO's science operations are currently confirmed until end of 2025. At that time, the mission-long data sets from these instruments will span 30 years and $\sim2.5$ solar cycles.

SOHO was primarily designed to study (i) helioseismic oscillations; (ii) the heating of the solar corona; and (iii) the solar wind and its acceleration \citep{domingo1995}. SOHO's observations of these phenomena are not only obviously valuable in their own right, but also as high-cadence, high-spatial-resolution, extremely-long-baseline insights into analogous phenomena in other stars and stellar systems. If we translate SOHO's solar observations into Sun-as-a-star observations---observations taken as if the Sun were a distant point source, viewed from a fixed perspective---we can compare them directly to observations of other stars. 

Some of SOHO's data has already been used this way, to great success. For example, the hourly-cadence solar light curve measured by SOHO's Variability of solar IRradiance and Gravity Oscillations (VIRGO) instrument \citep{frohlich1995} has been used to interpret the light curves, pulsation spectra, and asteroseismic parameters of other solar-type stars. In the early years of the SOHO mission, before the launch of either CoRoT in 2006 or Kepler in 2009, the VIRGO data served as a test light curve to analyze and model photometric stellar activity signals and mitigate their interference in exoplanet transit surveys (see e.g. \citealt{lanza2003, aigrain2004}). In more recent years, now that SOHO's observations span multiple solar activity cycles, the VIRGO light curve has been used to study how the helioseismic frequency of maximum power, $\nu_{\mathrm{max},\odot}$, varies over the solar cycle, and how this variation contributes to systematic uncertainty in the stellar parameters inferred from asteroseismic scaling relations calibrated to the Sun \citep{howe2020}. 

As another example, \cite{meunier2010a} used the magnetograms taken by SOHO's Michelson Doppler Imager (MDI) instrument \citep{scherrer1995} to simulate the expected solar radial velocity (RV) signal over magnetic cycle 23, an $\sim11.4$-year baseline, by summing up the expected RV signals of the MDI-observed plages, network structures, and sunspots on the solar surface. 
This work provided significant physical insight on its own--that the inhibition of convective blueshift by magnetically active regions dominates over the photometric contributions of plages and sunspots to the solar RV signal---but their simulated RV time series have also been used to test subsequent models of the RV signatures of stellar activity in general (e.g. \citealt{aigrain2012}), estimate exoplanet detectability limits around active stars, and optimize RV observation strategies to mitigate stellar activity signals (e.g. \citealt{meunier2015}).

Beyond the VIRGO light curve and the MDI magnetograms, however, the SOHO archive contains a wealth of lesser-known (among astronomers) long-baseline solar observations. These observations could be of great utility for astronomers interested in, for example, the radiation environment and habitability of exoplanets around solar-like stars: they span multiple solar magnetic cycles, so they capture the Sun's behavior at both low and high activity states, and several of SOHO's instruments were specifically designed to observe high-energy solar phenomena in ways that are extremely difficult or impossible to replicate for other stars, such as photometric monitoring in short-wavelength bandpasses inaccessible from the ground, or in-situ solar wind measurements. Some of SOHO's archival data is already calibrated and ready for analysis--for example, the soft X-ray solar light curve taken by the solar wind monitoring experiment called the Charge, Element, and Isotope Analysis System Solar Extreme-ultraviolet Monitor (CELIAS/SEM, \citealt{hovestadt1995}), which is taken in a 1-500\AA\ bandpass at 15-second cadence and spans SOHO's whole lifetime. Other data, meanwhile, requires further reduction before it can be used for Sun-as-a-star studies.

\begin{figure}
\begin{center}
\includegraphics[width=0.48\textwidth]{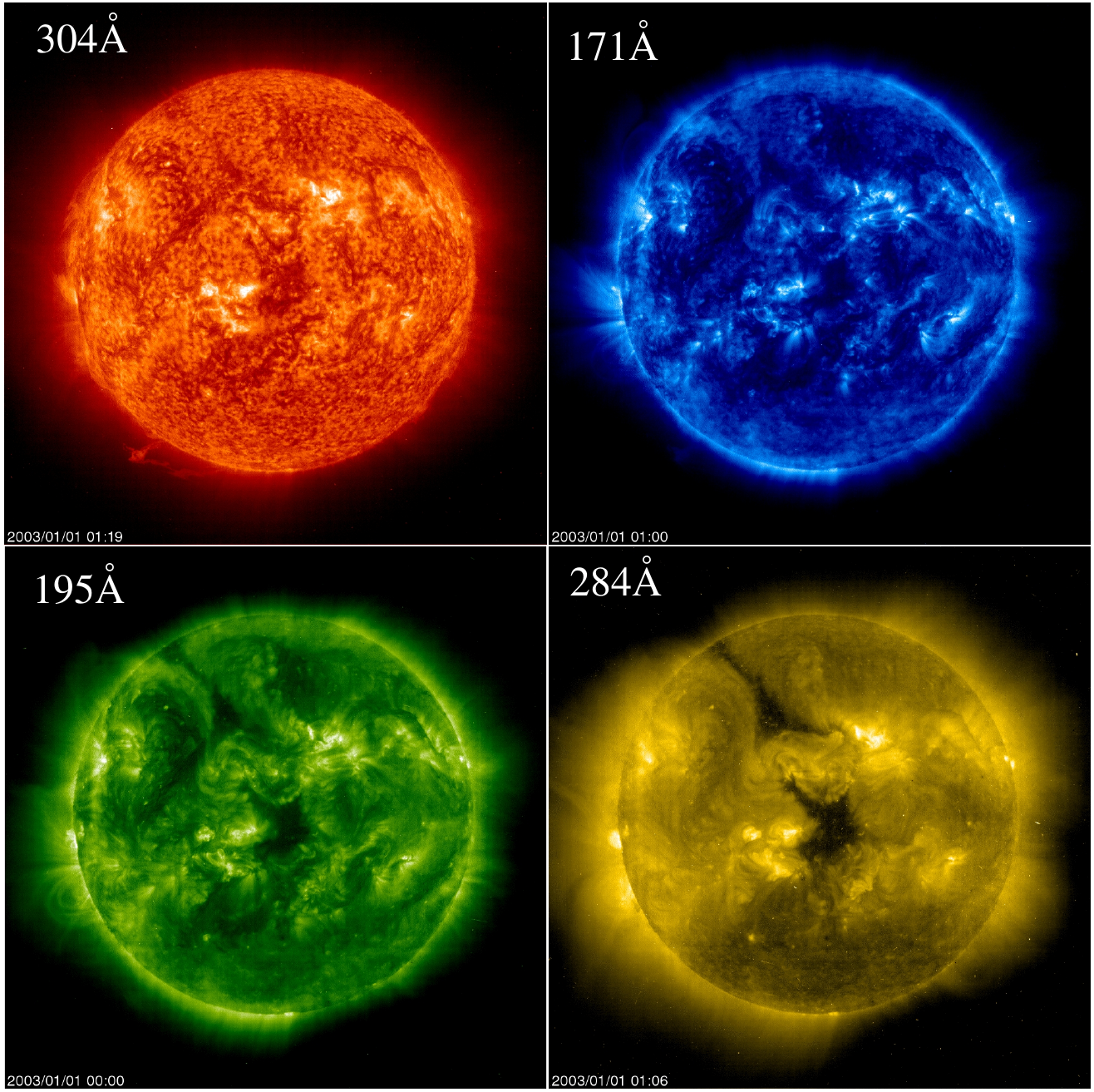}
\caption{Example images in EIT's four narrow UV bands, taken between 00:00 and 01:19 UTC on 1 January 2003. The intensity in each of these images is scaled logarithmically. The $304 $\AA\ band (upper left) observes the chromosphere and transition region, including HeII emission. The $171 $\AA\ (upper right), $195 $\AA\ (lower left), and $284 $\AA\ (lower right) bands observe progressively higher and hotter layers of the transition region and corona, designed to capture Fe IX-X, Fe XII, and Fe XV emission, respectively (\citealt{delaboudiniere1995}). In quiet periods, the 284\AA\ band is also contaminated by cooler material, because it overlaps with lines of Mg VII (see e.g. \citealt{young2023}).}
\label{fig:exampleEIT}
\end{center}
\end{figure}

In this paper, we present four light curves created from the full-disk solar images taken by SOHO's Extreme-ultraviolet Imaging Telescope (EIT, \citealt{delaboudiniere1995}). EIT images the solar disk, transition region, and corona, out to $1.5R_\odot$, in four narrow wavelength bands centered at $171 $\AA, $195 $\AA, $284 $\AA, and $304 $\AA, at a minimum cadence of once per day. These bands were chosen, in accordance with SOHO's second principal science goal, to capture the chromosphere, transition region, and solar corona. Figure~\ref{fig:exampleEIT} shows examples of EIT images in all four bands, and Figure~\ref{fig:EITbandpasses} shows the transmission curve associated with each bandpass. The 284\AA bandpass, in particular, has not been used in other EUV imagers so far except for NASA's Solar Terrestrial Relations Observatory (STEREO/EUVI), launched in 2006 \citep{STEREO}. This specificity adds to the uniqueness of the EIT database.



In Section~\ref{sec:data}, we explain the steps we took to reduce each EIT image into a single Sun-as-a-star light curve data point with a motivated uncertainty, accounting both for statistical uncertainty and mission-long instrumental systematics. In Section~\ref{sec:results}, we present the final EIT light curves, which are available for download from \href{https://doi.org/10.5281/zenodo.15474179}{https://doi.org/10.5281/zenodo.15474179}, and compare the EIT light curves to other long-baseline solar photometry from SOHO and from NASA's Solar Radiation and Climate Experiment Extreme-ultraviolet Photometer System (SORCE/XPS; \citealt{woods2005}), an independent space mission on a different orbit from SOHO which operated between 2003 and 2020. We conclude in Section~\ref{sec:discussion} with a discussion of the utility of these light curves for stellar and exoplanet astronomy.


\section{Data reduction}
\label{sec:data}
\subsection{SOHO's orbit and mission history}\label{subsec:history}

SOHO is on an elliptical halo orbit around the first Lagrange point in the Earth-Sun system, with a period of 178 days, or almost exactly 6 months \citep{ESAbulletin88}. Since launch in 1995, both the SOHO spacecraft and the EIT (see below) have experienced both slow degradation and acute hardware failures of varying severity. We must understand their effect on the EIT data, compensate for them where possible, and account for them in our uncertainty estimation. In Table~\ref{tab:datasummary}, we enumerate the various causes of missing or bad EIT data addressed in this section.

\begin{table*}
\caption{Reasons for missing or unusable EIT images} 
\centering 
\begin{tabular}{l l l l}
\hline\hline 
Reason & Affected bands & Effect & Text reference \\ [0.5ex] 
\hline 
 Near mission loss & All & No EIT images between 25 June and 12 Oct 1998 & ~\ref{subsec:history} \\
 High gain antenna failure & All & \makecell{Data gaps (``keyhole'' periods) of $\sim$ 2 weeks every \\$\sim$ 6 months between May 2003 and August 2009. \\Note  that these gaps overlap with  CCD bakeout \\periods, which happen throughout the mission \\lifetime.}& ~\ref{subsec:history}\\ 
 CCD bakeouts & All & \makecell{Data gaps of varying duration ($\sim$ a few days to \\$\sim$ 2 weeks) and frequency, often scheduled to \\coincide with keyholes. The combined keyhole-\\bakeout duty cycle is $\lesssim 15\%$ of the mission\\ baseline. Bakeouts also introduce discontinuities\\ in the EIT light curves.}  & ~\ref{subsec:history}, ~\ref{subsubsec:bakeouts} \\
 \makecell{Missing or corrupted\\ \ data blocks}& All& \makecell{Some images have missing pixels. $\sim9\%$ of the total \\number of EIT images must be discarded.}& ~\ref{subsec:EItprep}\\
 Pinholes & \makecell{284\AA\ band to\\ a significant extent; \\other bands negligibly}& \makecell{Before May 1998, a significant number of 284\AA\ \\images, and a handful of images in the other bands,\\ have bright spots due to pinholes in EIT's front heat-\\rejection filter. $5.6\% $ of the total number of 284\AA\ \\images must be discarded, and there is no 284\AA\ data\\ between August 1996 and May 1998.} & ~\ref{subsec:badpoints}\\
 Solar proton events & All & \makecell{$\sim5\%$ of all images appear ``staticky'' and must be \\discarded.}& ~\ref{subsec:badpoints}\\
 \makecell{Other one-off \\ \ camera errors} & All & \makecell{$<0.1\%$ of all images must be discarded due to random\\ problems (blurriness, underexposure, overexposure, etc.).}& ~\ref{subsec:badpoints}\\
\hline\hline 
\end{tabular}
\label{tab:datasummary} 
\end{table*}

Let us begin with acute hardware failures. Relevant to our analysis are two major events in SOHO's history: the near-total mission loss between June and October 1998 (see  \citealt{ESAbulletin97} for a full historical account), and the high-gain antenna pointing mechanism failure in 2003 \citep{ESAbulletin115}. 

The near-mission loss occurred on 25 June 1998, when a pair of interacting software errors caused the spacecraft to lose attitude control and then (because the antennae and solar panels were now pointing the wrong way) telemetry and power. Over four months of careful and creative effort, including using the Arecibo radio telescope and the NASA Deep Space Network 70-meter antenna to find the spacecraft, the SOHO team managed to re-establish contact, re-point toward the Sun, and eventually restore all twelve science instruments to full function. As a result of this event, there were no EIT images taken between 25 June and 12 October 1998, inclusive.

The high-gain antenna pointing mechanism began to fail in May 2003. By June, the SOHO team decided to stop trying to move the antenna altogether and instead leave it stationary at an angle where, for the majority of SOHO's  orbit, it can point toward Earth accurately enough for data transmission. This configuration necessitates that SOHO rotate by $180^{\circ}$ every half-orbit (i.e., every $\sim 3$ months) to keep the antenna pointed toward Earth while SOHO remains pointed at the Sun. Even so, twice per orbit, for approximately two weeks at a time (so-called ``keyhole'' periods), the high-gain antenna cannot be used, and the low-gain antenna must be used instead. Since data transmission through this antenna is slower, SOHO prioritizes data transmission from its helioseismic instruments (VIRGO; Global Oscillations at Low Frequencies, or GOLF; and the Michelson Doppler Imager, or MDI), which depend strongly on uninterrupted time series to achieve their scientific aims, during keyholes. For several years, this meant that no EIT images were taken during these periods. 

On 14 August 2009, the HGA position was improved such that EIT operations would  no longer be significantly interrupted by keyholes (J.B. Gurman and Jean-Philippe Olive, private comm.), although they remain interrupted every $3-6$ months by scheduled instrumental maintenance periods known as CCD ``bakeouts''. Starting from 2003, bakeouts were scheduled to deliberately fall during keyhole periods, and they have continued to be scheduled at roughly the same pace even now that keyholes are no longer a problem (see section~\ref{subsubsec:bakeouts}, below). The keyhole-bakeout duty cycle is $\lesssim 28$ days per $\sim 178$-day orbit, or $\lesssim 15\%$. The periodicity of the keyhole-bakeout events also has a strong effect on the periodogram of the final EIT light curves (see Section~\ref{sec:results}). 

\subsection{EIT images}\label{subsec:EItprep}

To begin, we downloaded the full catalog of EIT images ($\sim2$ terabytes) from ESA's SOHO science archive.\footnote{\url{https://ssa.esac.esa.int/ssa}} This catalog includes calibrated images (data processing level L1) from shortly after SOHO's launch in December 1995 to 31 December 2022, inclusive. Calibrated images from 1 January 2023 to 12 April 2024, inclusive,  were provided by Fr\'{e}d\'{e}ric Auch\`{e}re.\footnote{At time of resubmission, these later images are in process of being ingested into the ESA archive as well, and will hopefully be there by time of publication.} The methods described in this section will be equally applicable to later images. We present example images in EIT's four EUV passbands in Figure~\ref{fig:exampleEIT}. For querying and viewing EIT images from specific dates, or movies constructed from series of EIT images across a range of dates, see the SOHO movie theater\footnote{\url{https://soho.nascom.nasa.gov/data/Theater/}} and \texttt{JHelioviewer} application \citep{muller2017}.

These L1 calibrated EIT images used in this study have been pre-processed with the IDL routine \texttt{EITprep.pro}, available as part of the \texttt{SolarSoftWare} package \citep{freeland1998}. This routine:
\begin{enumerate}
    \item subtracts the analog-to-digital converter offset, which is measured by taking zero-exposure-time dark frames periodically throughout the mission;
    \item ``de-grids'' the image, to approximately remove the shadows cast on the CCD by the fine wire mesh grids supporting the aperture filter and the filter wheel;
    \item flat-fields the image;
    \item normalizes the counts per second of exposure time to calculate a flux, in units of data numbers per second;
    \item  normalizes the flux to the open position in EIT's filter wheel (see section~\ref{subsec:badpoints}, below); and
    \item normalizes the flux to account for EIT's declining pixel response over SOHO's lifetime, using a correlation with the Mg II solar activity index \citep{heath1986}.
\end{enumerate}

These steps address many, but not all, of the instrumental systematics affecting the EIT data. For example, the flat field correction accounts for the in-flight degradation of the CCD. It is based on ratios of images of an on-board white light calibration lamp, which are converted to EUV flat-field maps using an empirical relationship. The method satisfactorily corrects the relative spatial response variations, but there are known residuals in the absolute values that create temporal trends in the integrated fluxes. We address systematics correction, including temporal trends, in section~\ref{subsec:systematics}, below.

We restrict our analysis to only images taken after SOHO's commissioning phase was officially complete, on 16 April 1996. We discard any image that is not taken at EIT's full resolution, $1024 \times 1024$ pixels ($\sim 3\%$ of the total). To transmit images back to Earth, the SOHO LASCO electronics box, which is shared between EIT and LASCO (the Large Angle and Spectrometric Coronagraph experiment), divides each image into $32 \times 32$-pixel ``data blocks'' and downlinks the data block by block; we also discard any image with missing or corrupted blocks ($\sim 9\%$ of the total). 

The unit of flux in each EIT image pixel is counts per second; in other words, the images and the resulting light curves tally relative fluxes rather than absolute fluxes in physical units. This is the same as e.g. Kepler light curves, but different from e.g. the VIRGO total solar intensity time series, which are in physical units of watts per square meter. 

\begin{figure}
\begin{center}
\includegraphics[width=0.48\textwidth]{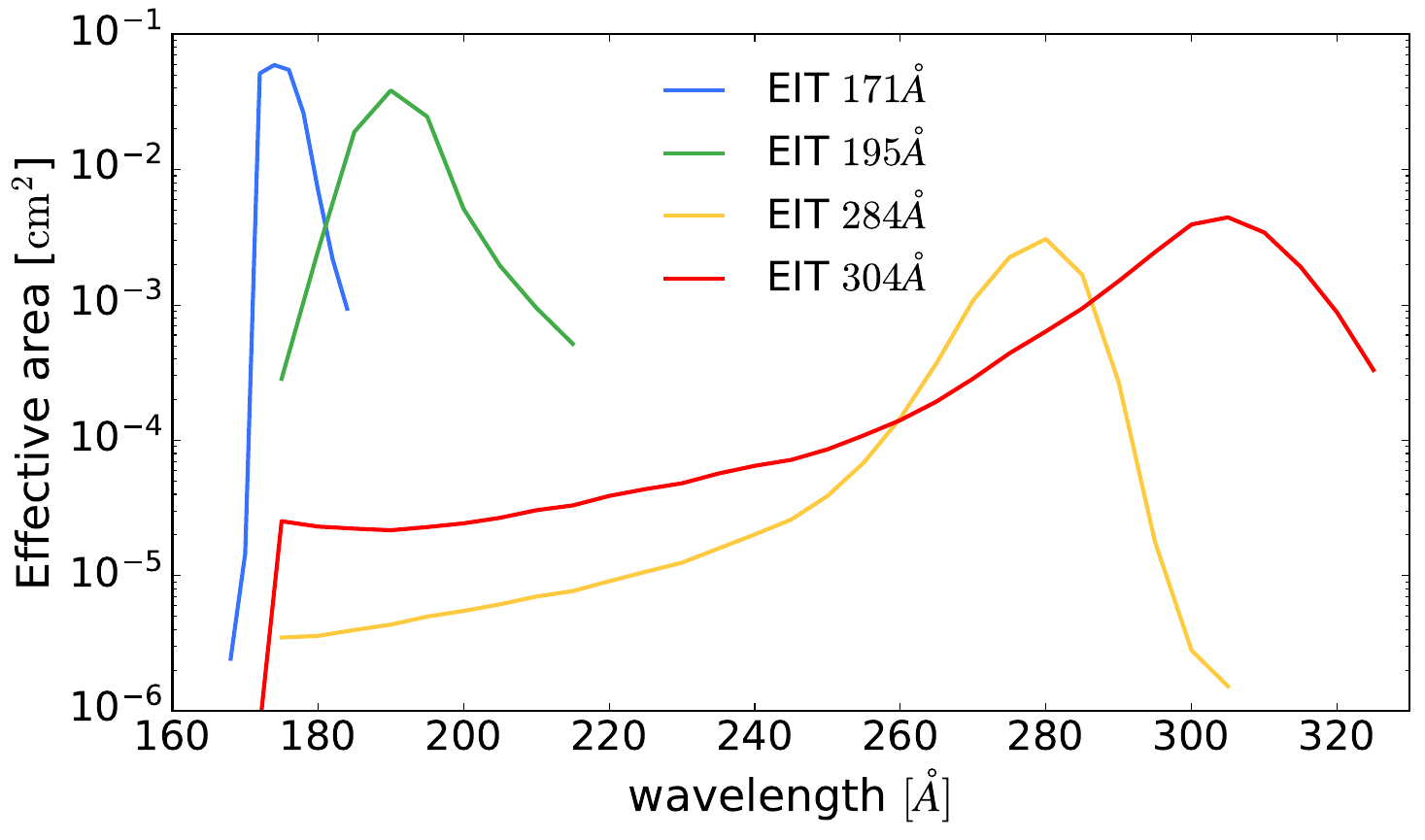}
\caption{The throughput of the four narrow EIT image bandpasses \citep{dere2000}. The $171$\AA\ bandpass has the highest peak throughput, and the $284$\AA\ bandpass has the lowest.}
\label{fig:EITbandpasses}
\end{center}
\end{figure}

Figure~\ref{fig:EITbandpasses}, plotted from tables 6-9 of \cite{dere2000}, shows the throughput of each of EIT's four narrow bandpasses. The bandpasses are defined by multilayer coatings on the four quadrants of the primary and secondary mirrors, but the overall throughput of the instrument (in any of the bandpasses) also depends on the position of the telescope's filter wheel, located between the secondary mirror and the detector. The throughput plotted in Figure~\ref{fig:EITbandpasses} corresponds to the filter wheel being set to its open position, with no additional filter (see section~\ref{subsec:badpoints} below). Note that the peak throughput differs by more than an order of magnitude between the $171 $\AA\ and $284 $\AA\ bandpasses; the corresponding (unnormalized) light curves will exhibit the same difference.


\begin{figure}
\begin{center}
\includegraphics[width=0.48\textwidth]{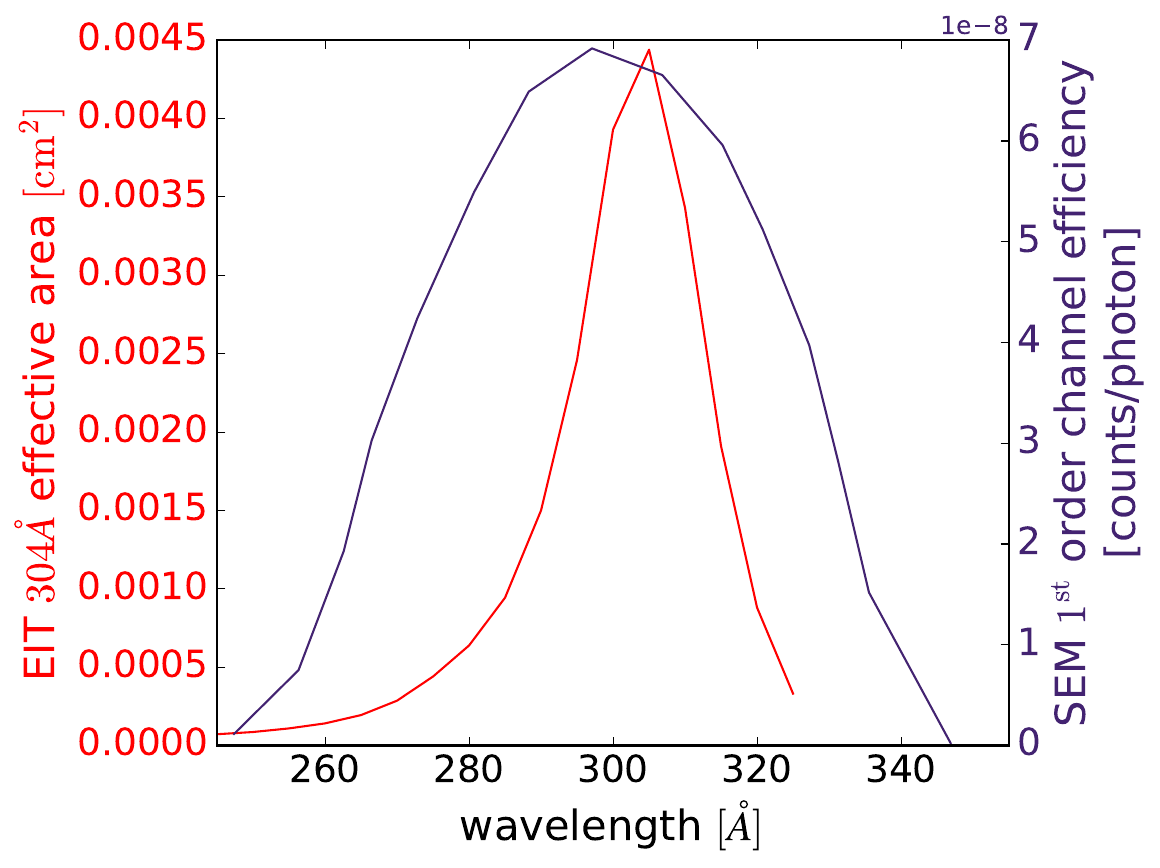}
\caption{A comparison of the throughput of the EIT $304$\AA\ bandpass \citep{dere2000} and the CELIAS/SEM $1^{\mathrm{st}}$-order channel bandpass, digitized from \citealt{hovestadt1995} Figure 23. Note that the y-axis in Figure~\ref{fig:EITbandpasses} is logarithmic, while the y-axis here is linear.}
\label{fig:EITSEMbands}
\end{center}
\end{figure}

\begin{figure}
\begin{center}
\includegraphics[width=0.48\textwidth]{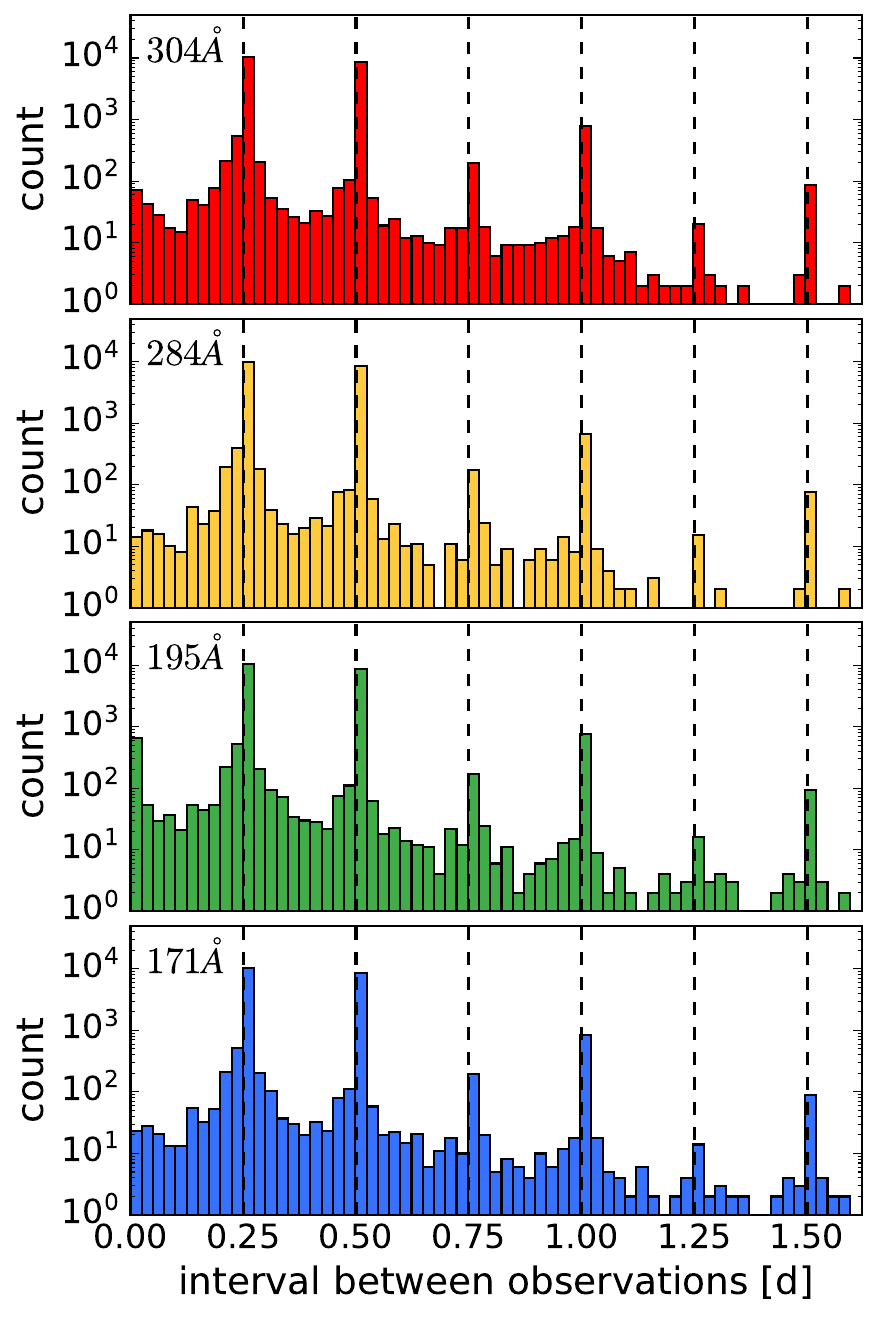}
\caption{Histograms of the intervals between successive EIT images taken in each of the four bandpasses. The modal cadence in all four filters is 0.25 days, but other multiples of 6 hours are also common.}
\label{fig:cadence}
\end{center}
\end{figure}

A histogram of the cadence of EIT observations in each bandpass is plotted in Figure~\ref{fig:cadence}. In all four bands, the modal cadence of the observations is 0.25 days for the first $\sim$ half of the mission, from 16 April 1996 to 20 July 2010. After this, the modal cadence decreases to 0.5 days. Other multiples of 6 hours are also common throughout the mission. The $195$\AA\ bandpass has an additional peak at extremely short cadences ($\sim 10$ minutes) because of a short-cadence observational campaign in this bandpass between 21 and 28 December 1996. 

\subsection{Benchmark solar time series for comparison}\label{subsec:otherdata}


We also download several time series from other solar instruments to compare to our EIT light curves and double-check our systematics corrections. The first of these is the VIRGO total solar irradiance time series, re-calibrated by \cite{finsterle2021}, which runs from February 1996 to December 2023.\footnote{Available at \href{ftp://ftp.pmodwrc.ch/pub/data/irradiance/virgo/TSI/}{ftp://ftp.pmodwrc.ch/pub/data/irradiance/virgo/TSI/}.} We use the TSI VIRGO/PMO6V-A+B time series (column name \texttt{TSI\_virgo\_fused\_new}), which is a composite time series produced by combining the data from VIRGO's two identical PMO6-V radiometers (PMO6-VA is used continuously, while PMO6-VB is a backup radiometer used to monitor PMO6-VA's degradation over time), and its associated uncertainty (column name \texttt{TSI\_virgo\_fused\_unc}).

For a long-baseline solar time series with ultraviolet spectral information, we download the mission-long NASA SORCE/XPS spectral irradiance time series \citep{woods2005}, also from LISIRD.\footnote{\href{https://lasp.colorado.edu/lisird/data/sorce_ssi_l3}{https://lasp.colorado.edu/lisird/data/sorce\_ssi\_l3}} This data set runs from 25 February 2003 to 25 February 2020, with a gap in observations from August 2013 to February 2014 due to battery degradation \citep{woods2021}. This data has been extensively calibrated against overlapping observations of the solar spectral irradiance by other instruments, so we will assume that it is not affected by any significant instrumental systematics.

Lastly, we download the version 4 CELIAS/SEM mission-long daily average time series \citep{hovestadt1995}, which we obtain from the LISIRD database.\footnote{\href{https://lasp.colorado.edu/lisird/data/soho_sem_P1D}{https://lasp.colorado.edu/lisird/data/soho\_sem\_P1D}} CELIAS/SEM has two bandpasses; the $0^{\mathrm{th}}$-order channel is a broadband soft X-ray to EUV bandpass, ranging from 1-500\AA. The $1^{\mathrm{st}}$-order channel is a narrow bandpass whose response function peaks at an almost identical wavelength to EIT's $304$\AA-band response function, as shown in Figure~\ref{fig:EITSEMbands}. Because of this similarity, we hypothesize that the normalized EIT $304$\AA\ light curve should be essentially identical to the normalized CELIAS/SEM $1^{\mathrm{st}}$-order light curve, as long as the respective instrumental systematics have been appropriately corrected. 

\begin{figure}
\begin{center}
\includegraphics[width=0.48\textwidth]{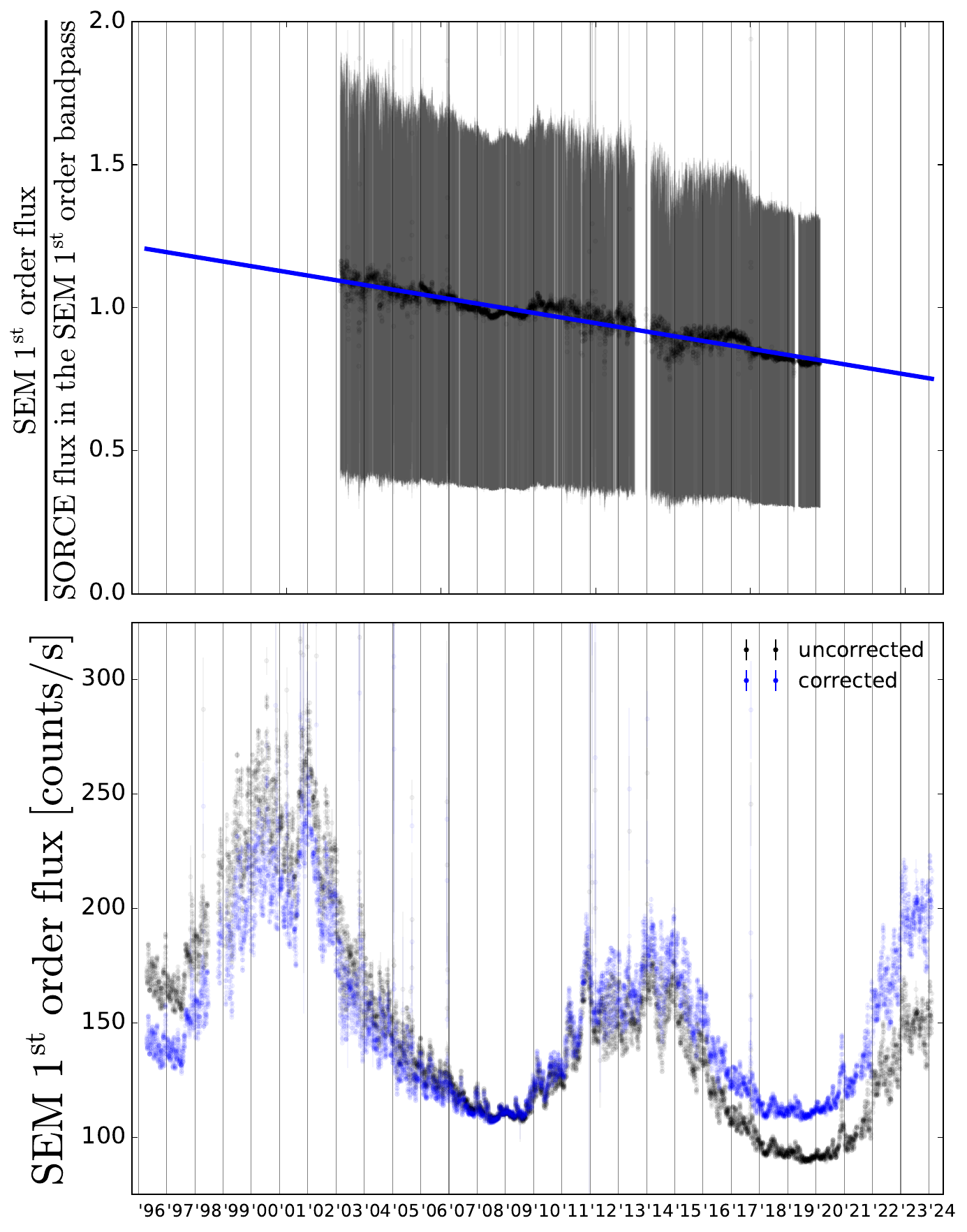}
\caption{Top panel: The ratio of the CELIAS/SEM $1^{\mathrm{st}}$-order light curve to a ``light curve'' constructed by propagating the SORCE/XPS solar spectral irradiance through the CELIAS/SEM $1^{\mathrm{st}}$-order bandpass. We assume that, since SORCE/XPS is well-calibrated against independent measurements of the solar spectral irradiance by other instruments, that the decline in this plot is evidence of long-term degradation in the CELIAS-SEM light curve, and fit a line (blue) with linear least squares fitting to account for it. Bottom panel: The uncorrected (black) and corrected (blue) versions of the CELIAS-SEM $1^{st}$-order light curve.}
\label{fig:SEMcorrection}
\end{center}
\end{figure}

We will assume that the CELIAS/SEM instrumental systematics have been appropriately corrected over short timescales. However, the LASP SOHO/SEM Version 4 Science Product release notes \citep{woodraska} say that ``there is a known issue with SEM data showing what appears to be uncorrected long-term degradation'', identified by comparing the CELIAS/SEM observations to ALMA observations. We can observe this degradation ourselves in Figure~\ref{fig:SEMcorrection} (upper planel), where we plot the ratio of the CELIAS/SEM $1^{\mathrm{st}}$-order light curve to a simulated ``light curve'' generated by propagating the SORCE/XPS spectral irradiance through the CELIAS/SEM $1^{\mathrm{st}}$-order bandpass. This ratio exhibits a linear decline over the SORCE/XPS baseline (2003-2020).

We fit the ratio with a line, using linear least squares fitting to find the optimal parameters, and then extrapolate the linear fit to cover the entire CELIAS/SEM baseline (1996-2024). We divide the extrapolated line out of the CELIAS/SEM $1^{\mathrm{st}}$-order light curve to produce a corrected version (Figure~\ref{fig:SEMcorrection}, lower panel). We propagate the analytic uncertainties on the line parameters from the least squares fit analytically through to the corrected light curve.

We will make extensive use of the corrected CELIAS/SEM $1^{\mathrm{st}}$-order light curve in our EIT systematics corrections (see section~\ref{subsec:systematics}, below). Before using it, we also perform a simple outlier rejection: we compute a running median of kernel size 11 for the light curve, then discard any point that is $< 0.75$ or $>1.25$ times the value of the running median. 


\subsection{Reducing each EIT image to a light curve data point}\label{subsec:reduction}

To convert an EIT image into a single light curve flux value with an uncertainty, we must account for (i) the flux coming from the solar disk and (ii) the coronal flux coming from beyond the solar limb. As Figure~\ref{fig:exampleEIT} shows, the off-disk contribution is relatively minor in the $304$\AA\ band, which observes the chromosphere and transition region, but increasingly important for the $171$\AA, $195$\AA , and $284$\AA\ bands, which observes higher layers of the corona. (Note that the intensity in each image in  Figure~\ref{fig:exampleEIT} scales logarithmically.)

The on-disk flux is easy to sum up, because the solar disk is fully contained within the bounds of every EIT image. The off-disk flux requires slightly more care, because the Solar disk subtends a different angular diameter in each image depending on SOHO's distance from the Sun. When SOHO is closer to the Sun, the Sun occupies a larger fraction of EIT's field of view, and vice versa. 

We correct for this effect by normalizing EIT's field of view to that of the EIT image taken at minimum distance from the Sun across the whole EIT baseline (i.e., the image where the Solar disk occupies the greatest fraction of the field of view). We calculate the sky-projected physical distance spanned by that image, then (in effect) ``trim'' the outermost row(s) of pixels off every other image to create a square of the same physical size.

EIT's angular field of view is:

\begin{equation}
    \mathrm{FOV} = 1024 \mathrm{pix} * 2.627\mathrm{''/pix}
\end{equation}

where 2.627 arcseconds per pixel is the pixel scale, which is the same for every EIT image in both the $x$ and $y$-directions \citep{auchere2000}.

When SOHO is at its minimum distance from the Sun, $D_\mathrm{min}$, the sky-projected physical distance spanned by the image is

\begin{equation}
    p = 2 D_{\mathrm{min}} \tan{\left(\frac{\mathrm{FOV}}{2}\right)}
\end{equation}

The angle that subtends the same physical distance $p$ when SOHO is at an arbitrary distance $D$ from the Sun is

\begin{equation}
    \delta = 2 \arctan{\left(\frac{p}{D}\right)}
\end{equation}

and the corresponding diameter of the image, in units of pixels, is

\begin{equation}
    d_{\mathrm{pix}} = \mathrm{round}\left(\frac{\delta}{2.627}\right),
\end{equation}

where $\delta$ is expressed in arcseconds. $d_{\mathrm{pix}}$ varies from 988 pixels at SOHO's maximum heliocentric distance to 1024 pixels at SOHO's closest approach.

\begin{figure*}
\begin{center}
\includegraphics[width=\textwidth]{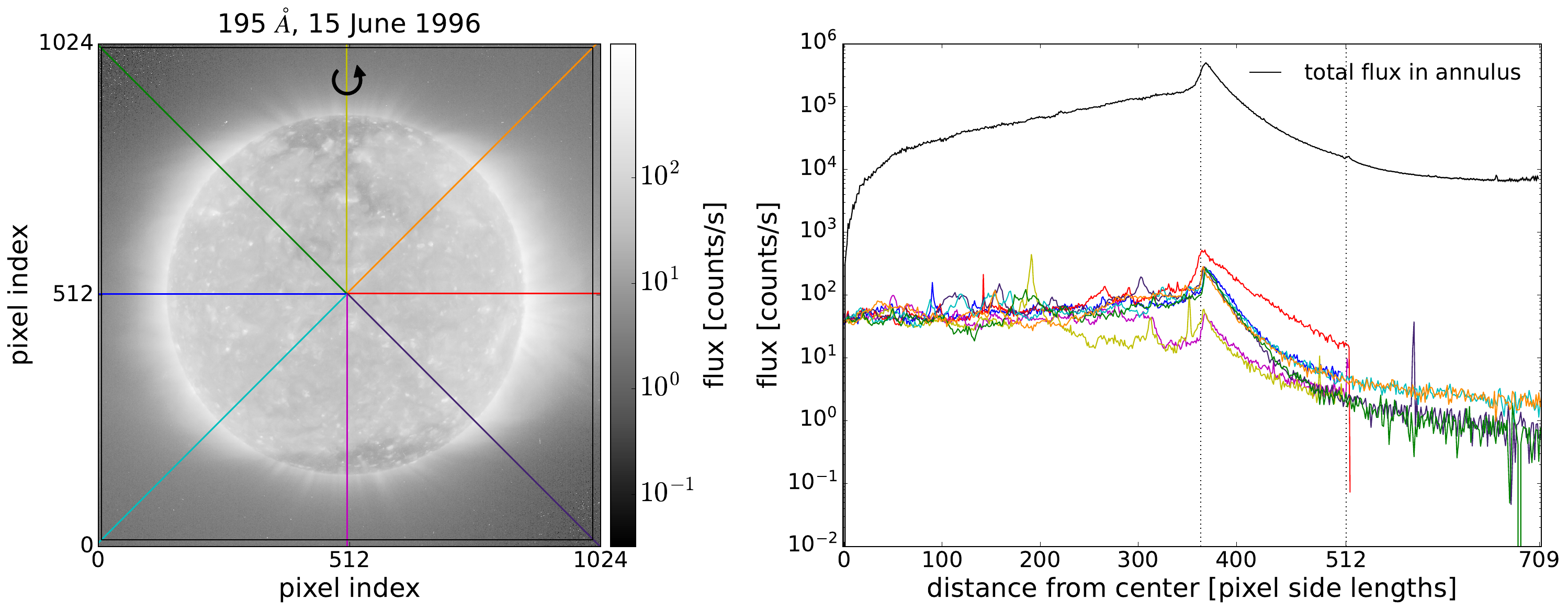}
\caption{Left panel: An example EIT 195\AA\ image taken on 15 June 1996. The curved arrow shows the direction of the solar rotation. The eight bisecting lines meet at the center of the solar disk, $(i_{\mathrm{center}}, j_{\mathrm{center}})$. The image is color-mapped by the logarithmic flux value in each pixel in units of counts per second. The black square just inside the border of the image is the ``trimmed'' square of side length $2r_{\mathrm{pix}}$ centered at the center of the solar disk. Right panel, colored lines: The flux value per pixel along each example bisecting line, as a function of distance from $(i_{\mathrm{center}}, j_{\mathrm{center}})$. Note that, beyond the solar limb at $r \simeq 363$ (the leftmost vertical dashed line indicates the photospheric radius), the lines that intersect the solar poles (yellow, magenta) have much lower average flux per pixel than the other lines, due to significantly fainter coronal emission from a polar coronal hole. Black line with (barely visible) error bars: The total flux in each annulus from radius $r = 1$ to $r = r_\mathrm{pix}\sqrt{2} = 709$ pixels and its associated uncertainty. The vertical dashed line at $r=512$ indicates where annuli begin to overlap the edge of the image and our flux extrapolation begins.}
\label{fig:fluxAnnuli}
\end{center}
\end{figure*}

We proceed by summing up the flux in each image, out to its normalized diameter $d_{\mathrm{pix}}$. Figure~\ref{fig:fluxAnnuli} shows our approach to tallying the on- and off-disk flux in each image graphically. Briefly, in order to include as much off-disk flux as possible, we extrapolate beyond the bounds of the image to include an estimate of all the flux within a circle of radius $\sim r_\mathrm{pix}\sqrt{2}$ pixels of the center of the Sun, where $r_{\mathrm{pix}}$ = $d_{\mathrm{pix}}/2$. In other words, $r_{\mathrm{pix}}$ is the distance between the center of the Sun and any corner of the image, assuming the Sun were perfectly centered. 

To do this extrapolation, we divide the image into narrow annuli centered at the center of the solar disk, and sum up the flux contribution from each annulus. Annuli with outer radii larger than $r_{\mathrm{pix}}$ pixels overlap with the edge of the image, so for these annuli, we linearly extrapolate the flux from the observed segment of the annulus to calculate the flux expected from the full annulus. 

For each EIT image, we begin by calculating the $x$ and $y$ indices of the pixel closest to the center of the solar disk, using metadata from the FITS header. Every EIT image has the same reference pixel, located at the center of the CCD, at position $(i_{\mathrm{ref}},j_{\mathrm{ref}}) = (512.5, 512.5)$ pixels (recorded in header keywords CRPIX1 and CRPIX2). 

However, the $x$-axis of the CCD array is not necessarily aligned with the solar equator, and the center of the CCD is not necessarily aligned with the center of the Sun. The header keyword CROTA gives the rotation angle $\theta_{\mathrm{rot}}$ (in degrees) between the CCD $x$-axis and the solar $x$-axis (in helioprojective cartesian coordinates), defined by the solar equator. The header keywords CRVAL1 and CRVAL2 give the helioprojective cartesian coordinates of the reference pixel of the CCD (in arcseconds).

We proceed by transforming the position of the reference pixel (CRVAL1, CRVAL2) to the CCD coordinate system:

\begin{gather}
 \begin{bmatrix} x_{\mathrm{ref}} \\ y_{\mathrm{ref}} \end{bmatrix}
 =
  \begin{bmatrix}
   \cos{\theta_{\mathrm{rot}}} &
   \sin{\theta_{\mathrm{rot}}} \\
   -\sin{\theta_{\mathrm{rot}}} &
   \cos{\theta_{\mathrm{rot}}} 
   \end{bmatrix}
   \begin{bmatrix}
   \mathrm{CRVAL1} \\
   \mathrm{CRVAL2}
   \end{bmatrix}
\end{gather}

By converting the vector $(x_{\mathrm{ref}}, y_{\mathrm{ref}})$ to units of pixels and following it backwards from $(i_{\mathrm{ref}}, j_{\mathrm{ref}})$, we can find the CCD coordinates of the origin of the helioprojective coordinate system, i.e., the solar center:

 \begin{align}
i_{\mathrm{center}} &= i_{\mathrm{ref}} - \frac{x_\mathrm{ref}}{2.627} \nonumber \\
j_{\mathrm{center}} &= j_{\mathrm{ref}} - \frac{y_\mathrm{ref}}{2.627}.
 \end{align}

Most EIT images are near-centered on the Sun, with $(i_{\mathrm{center}}, j_{\mathrm{center}}) \simeq (508, 518)$. In the left panel of Figure~\ref{fig:fluxAnnuli}, the bisecting lines meet at $(i_{\mathrm{center}}, j_{\mathrm{center}})$.

We proceed by dividing the image into annuli of width 1 pixel, centered at $(i_{\mathrm{center}}, j_{\mathrm{center}})$, stepping outward in radius from $r = 1$ pixel to $r =  \mathrm{round}(r_\mathrm{pix}\sqrt{2})$ pixels. Pixels with indices $(i, j)$ such that 

\begin{equation}
    (r - \frac{1}{2})^2 < [(i - i_{\mathrm{center}})^2 + (j - j_{\mathrm{center}})^2] \leq (r + \frac{1}{2})^2
\end{equation}

belong to the annulus of radius $r$.

For $r \lesssim 512$, the entire annulus is contained within the boundaries of the square image, so the total flux in the annulus is
\begin{equation}
    F_r = \sum_{k=1}^N F_k,
\end{equation}

where the sum is over all $N$ pixels in annulus $r$. We approximate the uncertainty on the total flux in the annulus as Poisson uncertainty, given by
\begin{equation}\label{eq:annUnc}
    \sigma_{F_r} \simeq \sqrt{F_r}.
\end{equation}

We note that this estimate formally underestimates the appropriate uncertainty on $F_r$, because it does not include for thermal signal in the CCD; however, given that the CCD is operated at a temperature of $\sim -70^{\circ}$C, we expect the thermal signal to be negligible compared to the solar signal.


For $r \gtrsim  512$, the annulus overlaps the edge of the image, and we must extrapolate beyond the known pixels to estimate the total flux in the annulus. We do this by calculating the total flux value over the $N_{\mathrm{known}}$ observed pixels in the annulus,
\begin{equation}
    F_{\mathrm{known}} = \sum_{k=1}^{N_{\mathrm{known}}} F_k,
\end{equation}

and then linearly scaling this known flux to the extrapolated number of pixels that would be in the annulus if the image were bigger, $N_{\mathrm{extrapolated}} = 2\pi r$. The total flux in the annulus is then
\begin{equation}
    F_r = F_{\mathrm{known}} \frac{N_{\mathrm{extrapolated}}}{N_{\mathrm{known}}}, 
\end{equation}

and its uncertainty, propagated analytically, assuming uncorrelated noise, is
\begin{align*}
    \sigma_{F_r} &= \sqrt{ \left(\frac{\partial F_r}{\partial F_{\mathrm{known}}}\right)^2 \sigma_{F_{\mathrm{known}}}^2 +  \left(\frac{\partial F_r}{\partial N_{\mathrm{extrapolated}}}\right)^2 \sigma_{N_{\mathrm{extrapolated}}}^2} \\
    &= \sqrt{ \left( \frac{N_{\mathrm{extrapolated}}}{N_{\mathrm{known}}} \right)^2 F_{\mathrm{known}} +  \left( \frac{F_{\mathrm{known}}}{N_{\mathrm{known}}} \right)^2 2 \pi r},
\end{align*}

assuming that $\sigma_{N_{\mathrm{extrapolated}}} = \sqrt{2\pi r}$. The right panel of Figure~\ref{fig:fluxAnnuli} shows the flux in each annulus as a function of its radius $r$ and its uncertainty. The uncertainties for the annuli overlapping the edge of the image are (appropriately) very large. 

We obtain the total flux in the image, $F_{\mathrm{image}}$, and its associated uncertainty by summing the annular fluxes in quadrature:

\begin{align*}
    F_{\mathrm{image}} = \sum_{r=1}^{r_{\mathrm{max}}} F_r\\
    \sigma_{F_{\mathrm{image}}} = \sqrt{\sum_{r=1}^{r_{\mathrm{max}}} \sigma_{F_r}^2},
\end{align*}

where $r_{\mathrm{max}}$ is the maximum distance, in units of pixels, between $(i_{\mathrm{center}}, j_{\mathrm{center}})$ and any corner of the image. 

The flux in each image is generally dominated by the on-disk flux. For the $284$\AA\ band, where the off-disk region is brightest, the off-disk flux contributes $\sim25-30\%$ of the total. The off-disk flux in the extrapolated region ($r \gtrsim 512$) is contributes at most $\sim5\%$ to the total.

\subsection{Identifying and discarding bad light curve points}\label{subsec:badpoints}

\begin{figure*}
\begin{center}
\includegraphics[width=\textwidth]{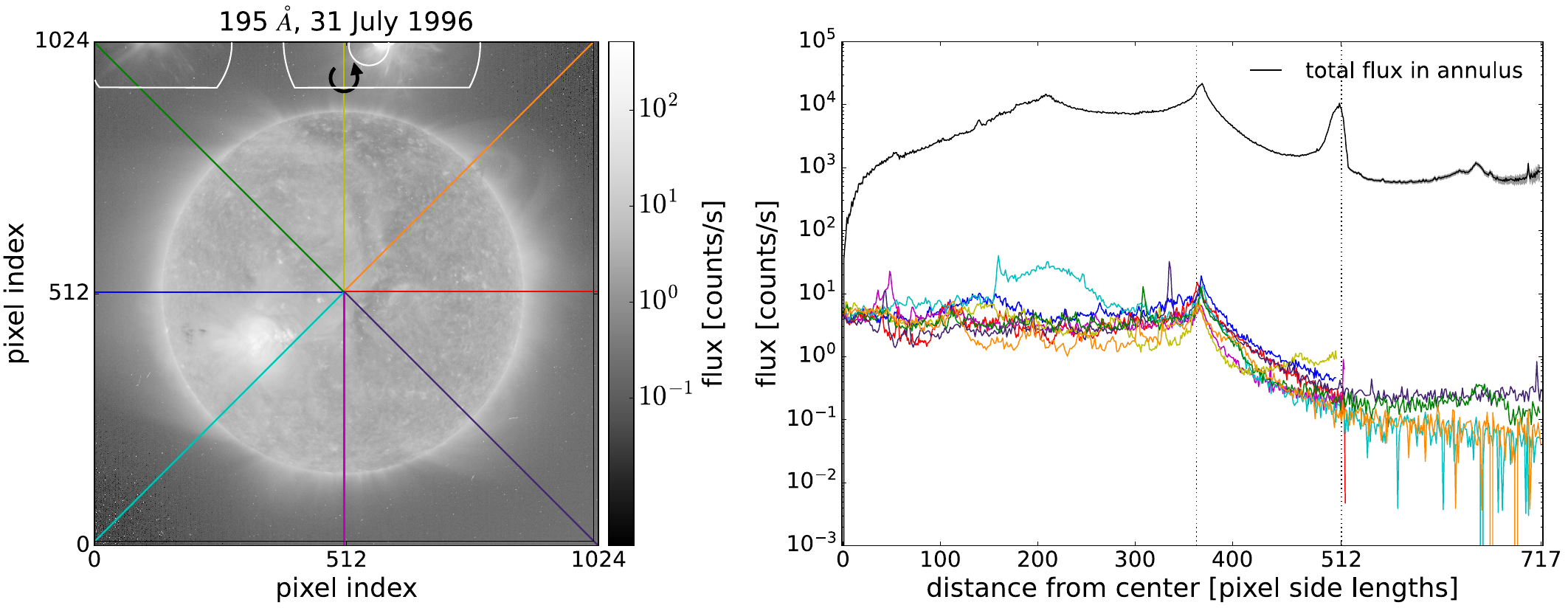}
\caption{Same as Figure~\ref{fig:fluxAnnuli}, but showing an example EIT $284$\AA\ image taken on 31 July 1996 with bright spots along its upper edge caused by pinholes in EIT's front filter. Note the significant bump in annular flux at $r \simeq 500$ due to the pinholes. The three patches we use to identify images with pinholes are shown with white outlines; only the two larger patches are necessary for this particular image, but the smaller patch is  useful for catching images in the 304\AA\ filter with smaller pinholes.}
\label{fig:pinholes}
\end{center}
\end{figure*}

After using the above procedure to create preliminary light curves in each of the four EIT bands, we next identify and discard individual light curve data points afflicted by instrumental problems. The first problem is that EIT's front aluminium heat-rejection filter (see \citealt{delaboudiniere1995}, section 2.1) developed small pinholes early in the mission, likely during launch. These pinholes allow excess light through to the optical cavity and cause bright spots to appear along the edges of some of the EIT images; see Figure~\ref{fig:pinholes} for an example. Due to the position of the pinholes, the $284 $\AA\ band is much more frequently affected than any other band.

Luckily, EIT was built with a redundant aluminium filter along its optical path. In the early years of the SOHO mission, this filter was not used, in order to keep exposure times as low as possible. However, during this time, the front filter began to peel away from its mounting, leading to a dramatic increase in leaked light onto the CCD, so in May 1998, the operations team decided to start using the redundant filter in order to block this light. The \texttt{EITprep.pro} routine (see section~\ref{subsec:EItprep}) normalizes the flux in each EIT image to the value it would have without this additional filter (i.e., normalizes it to the open position in the filter wheel), so images before and after May 1998 are still directly comparable.

Because the pinholes always appear in the same part of the image, it is easy to identify images that have them. We define three patches enclosing the worst pinhole spots, each of which is a circle cut off at the top by the edge of the image: one at $(i,j) = (557, 1016)$  of radius 40 pixels, one at $(i,j) = (130, 1022)$ of radius 150 pixels, and one at $(i,j) = (585, 1022)$ of radius 200 pixels. These latter two are also cut off below $j=931$ to avoid catching too much of the corona, which is actually (and not spuriously) bright. The three patches are plotted as white outlines in Figure~\ref{fig:pinholes}. 

For each pre-May 1998 image, we calculate the mean flux per pixel within each patch and compare it to the mean flux per pixel in a reference patch of the same size, shape, and $i$-index, reflected across the solar equator. If the mean flux per pixel in any of the three pinhole patches is more than twice the mean flux per pixel in its respective reference patch, we flag the image as containing a pinhole and discard it. We discard 17 out of 22645 $304$\AA\ images (0.08\%), 1276 out of 22673 $284$\AA\ images (5.6\%), 9 out of 23354 $195$\AA\ images (0.04\%), and 6 out of 22574 $171$\AA\ images (0.03\%) because of pinholes. The $284$\AA\ light curve, the worst affected, has no data between August 1996 and May 1998.

The second cause of bad data points is solar proton events (SPEs), outbursts of energetic protons associated with high-energy solar flares or coronal mass ejections. Solar proton events cause a static-like pattern across EIT's CCD and result in the CCD recording artificially low flux values. This static appears in images taken in all four EIT bands during the SPE. 

To identify and discard SPE-affected images, we download the NOAA Space Environment Services Center catalog of SPE events.\footnote{Available at \url{ftp://ftp.swpc.noaa.gov/pub/indices/SPE.txt}.} We consider the 141 SPEs in this catalog that occurred between the beginning of SOHO's post-commissioning operations on 16 April 1996 and the end of our current data set on 12 April 2024.

For each SPE, we calculate the ``rise time'' $t_{\mathrm{rise}}$ as the interval between its beginning and time of maximum $t_{\mathrm{max}}$, both given in the catalog. We then discard any EIT image taken within the interval $[t_{\mathrm{max}} - 2t_{\mathrm{rise}}, t_{\mathrm{max}} + 4t_{\mathrm{rise}}]$, which by inspection is sufficient to bracket the entire associated dip in EIT flux for all 141 SPEs. In total we discard 1045 images from the $304$\AA\ band, 1049 images from the $284$\AA\ band, 1046 images from the $195$\AA\ band, and 1052 images from the $171$\AA\ band due to SPEs.

After converting the images into preliminary light curves, we also eliminate and discard a handful of outlying data points which are either anomalously low or anomalously high in flux. We identify these points by the same procedure as the CELIAS/SEM light curve outier rejection: computing a running median of kernel size 11 for each light curve, then discarding any point that is $< 0.75$ or $>1.25$ times the value of the running median. Inspection of the images corresponding to these outliers suggests that they are due to one-off camera errors (i.e., most of these images appear blurry, underexposed, overexposed, or some combination). We discard 19 outliers from the 304\AA\ light curve, 19 from the 284\AA\ light curve, 28 from the 195\AA\ light curve, and 14 from the 171\AA\ light curve.

\subsection{Light curve-level systematics corrections}\label{subsec:systematics}

Before the preliminary EIT light curves can be treated as Sun-as-a-star observations, we must make several systematics corrections at the light curve level.

\subsubsection{Heliocentric distance correction}

First, SOHO's heliocentric distance varies significantly throughout its orbit. This has two effects: first, that the Sun appears correspondingly larger or smaller in EIT's field of view, and second, that the observed Solar flux varies according to the inverse square law. We correct for the first effect by trimming each image to match the physical dimensions of the image taken at SOHO's closest approach to the Sun. To correct for the second, we normalize each light curve data point by SOHO's heliocentric distance $d$ at the time of its observation, to a reference heliocentric distance of 1 AU:

\begin{equation}
    F_{\mathrm{dist.\ corrected}} = F \left(\frac{d}{1 \mathrm{AU}}\right)^2.
\end{equation}

\subsubsection{Correcting for 304\AA\ exponential degradation and CCD bakeouts}\label{subsubsec:bakeouts}
\begin{figure*}
\begin{center}
\includegraphics[width=\textwidth]{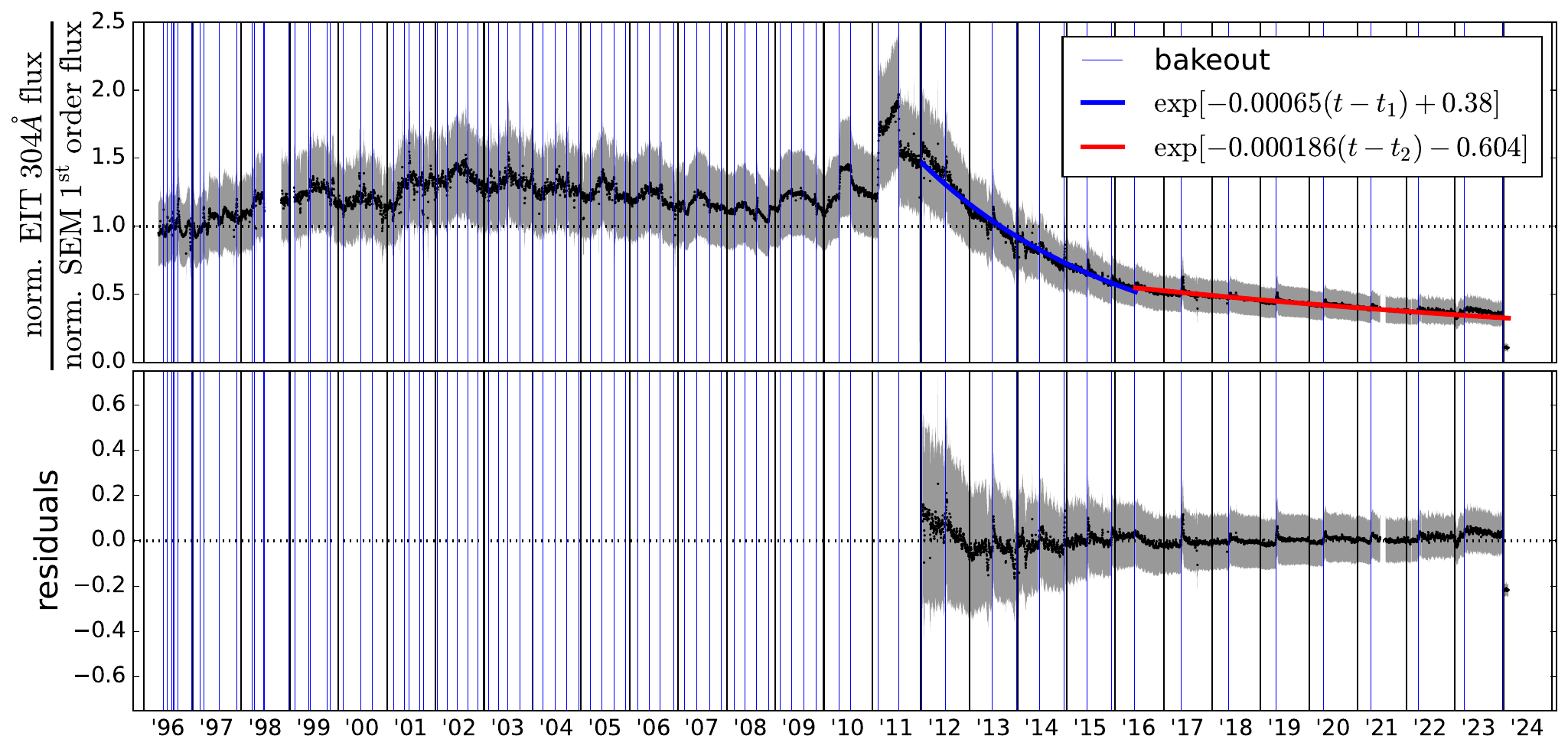}
\caption{Upper panel: The ratio of the normalized EIT $304$\AA\ light curve to the CELIAS/SEM $1^{\mathrm{st}}$-order light curve. The two exponential functions fit to the post-2012 dropoff are plotted in blue and red, respectively, with best-fit parameters given in the figure legend. $t_1$, the beginning of the first exponential trend, is 11 January 2012, and $t_2$, the beginning of the second exponential trend, is 3 June 2016.  Note that the SEM time series ends on 8 Feb 2024 (earlier than our EIT light curves), so this ratio plot also ends on 8 February 2024.  Lower panel: The residuals of the exponential fit.} 
\label{fig:exponential}
\end{center}
\end{figure*}

Next, we have to account for systematic changes to EIT's sensitivity, beyond those already divided out by \texttt{EITprep.pro}. We can identify these by examining the mission-long time series of the ratio of the normalized EIT $304$\AA\ light curve to the normalized CELIAS/SEM $1^{\mathrm{st}}$-order light curve, plotted in the upper panel of Figure~\ref{fig:exponential}. 

The vertical blue lines in this figure mark ``bakeout'' periods. EIT's optical cavity is not a perfect vacuum; there is a small amount of volatile material inside, probably water and hydrocarbons \citep{moses1997, gurman}. This material tends to condense onto the CCD, which is kept cold by passive radiative cooling through the shadowed side of the telescope, and block some EUV light. The EIT operations team periodically heats up, or ``bakes out'', the CCD to evaporate this material. Additionally, the bakeouts help maintain the CCD's performance by fixing ``electron traps:'' wells of trapped charge in random pixels caused by cosmic ray hits and EUV overexposure \citep{gurman}. 

As Figure~\ref{fig:exponential} shows, there are local trends in the EIT $304$\AA\ to CELIAS/SEM ratio specific to each inter-bakeout period, and also an approximately exponential decrease in this ratio from early 2012 onward. 

\begin{figure*}
\begin{center}
\includegraphics[width=\textwidth]{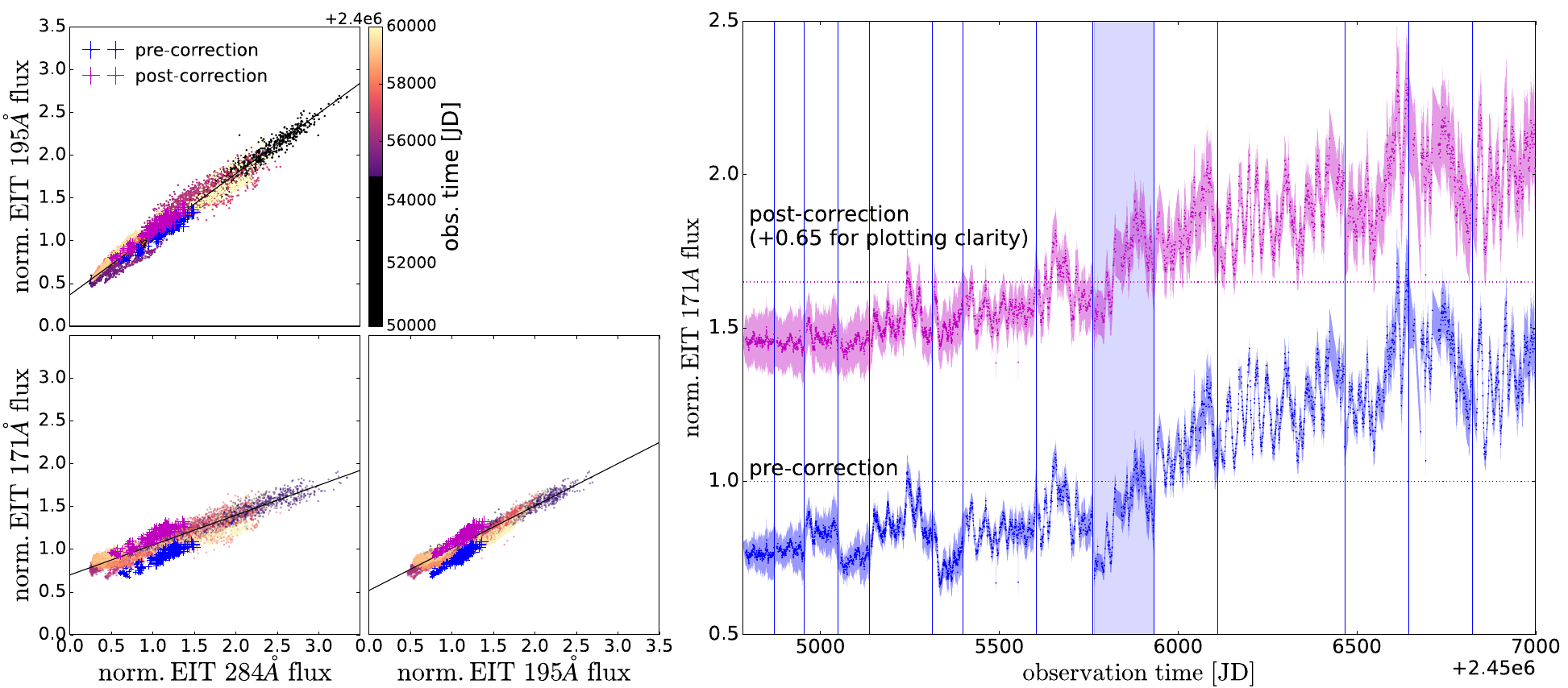}
\caption{Left: Scatterplots of the light curves in the EIT $171$\AA\, $195$\AA\, and $284$\AA\ bands, after the initial corrections to the post-12 November 2008 inter-bakeout segments are applied. All data points pre-12 November 2008 are plotted in black; the later points are color-mapped by observation time. The uncertainties of the points are not plotted, for clarity. Straight lines: Projections of the three-dimensional line of best fit to the pre-12 November 2008 data. Blue $+$ symbols: The inter-bakeout segment between 22 July 2011 and 6 January 2012. Magenta $+$ symbols: The same inter-bakeout segment, with additional constant offsets applied to move it closer to the line of best fit along the shortest possible vector. Right: A segment of the EIT 171\AA\ light curve, from 12 November 2008 to late 2014, when the bakeout discontinuities are most extreme. In blue is the light curve with only the initial bakeout correction applied; in magenta is the light curve with an additional constant offset added per bakeout. The magenta light curve is offset by +0.65 compared to the blue one for plotting clarity. The discontinuities across the bakeouts are much less extreme in the magenta version. The shaded region shows the inter-bakeout segment between 22 July 2011 and 6 January 2012, which is highlighted in the left panel.}
\label{fig:bakeoutLineup}
\end{center}
\end{figure*}

Curiously, while the local inter-bakeout trends affect all four EIT bands in a similar way, the late exponential dropoff appears to only affect the $304$\AA\ band. The evidence for this being $304$\AA\-specific behavior includes (i) that there is no exponential trend in the ratio of any of the other three bands to any independent instrument; and (ii) that the flux values in all four EIT bands have a first-order linear relationship to each other before 2012, but this relationship breaks down between the $304$\AA\ band and any other band from 2012 onward. Meanwhile, the linear relationships between the $284$\AA, $195$\AA, and $171$\AA\ bands persist throughout the mission (see Figure~\ref{fig:bakeoutLineup}). 

We hypothesize that this is due to the $304$\AA\ band's known high sensitivity to both molecular condensation onto the CCD and to radiation-induced degradation, the two problems that CCD bakeouts attempt to ameliorate, compared to the other bands. As can be seen from the vertical blue lines in Figure~\ref{fig:exponential}, the $304$\AA\ exponential trend begins relatively shortly after the decision to decrease the frequency of bakeouts, first to every six months (starting in mid-2010) and then to every year (starting in mid-2016). It is possible that these less frequent bakeouts are failing to fully compensate for the effects of condensation or electron traps on the $304$\AA\ images.

To proceed, therefore, we fit two exponential functions to the ratio of the EIT $304$\AA\ light curve to the CELIAS/SEM light curve, using simple linear least squares fitting, of the form:
\begin{equation}
    \ln{R} = A(t-t_{\mathrm{ref}}) + B,
\end{equation}

where $R$ is the ratio of the two light curves. The first exponential function is to the time interval between $t_1$ = 11 January 2012, or JD 2455938.417 (the end of the bakeout corresponding to the visible onset of the exponential trend) and 3 June 2016 (the end of the bakeout corresponding to the change from baking out every six months to every year), so $t_{\mathrm{ref}} = t_1$. The second is to the time interval between $t_{\mathrm{ref}} = t_2$ = 3 June 2016, or JD 2457542.583333, and the end of the time series. The best-fit exponential functions are plotted in blue and red, respectively, in the upper subplot of Figure~\ref{fig:exponential}, and the residuals left over in the ratio after these exponential trends are divided out are shown in the lower subplot of Figure~\ref{fig:exponential}. The best-fit parameters and their uncertainties are given in Table~\ref{tab:expParams}.

\begin{table}
\caption{Best-fit parameters of the two exponential functions fit to the ratio of the normalized EIT 304\AA\ light curve to the CELIAS/SEM $1^{\mathrm{st}}$-order light curve with linear least squares fitting.} 
\centering 
\begin{tabular}{l l}
\hline\hline 
Parameter & Best-fit value \\ [0.5ex] 
\hline 
  $A_1$ &  $-0.00065 \pm 0.00001 $\\
  $B_1$ &  $0.38 \pm 0.01$\\
  $A_2$ &  $-0.000186 \pm 0.000006 $\\
  $B_2$ &  $-0.604 \pm 0.009 $\\
  [1ex]
\hline\hline 
\end{tabular}
\label{tab:expParams} 
\end{table}

\begin{figure*}
\begin{center}
\includegraphics[width=\textwidth]{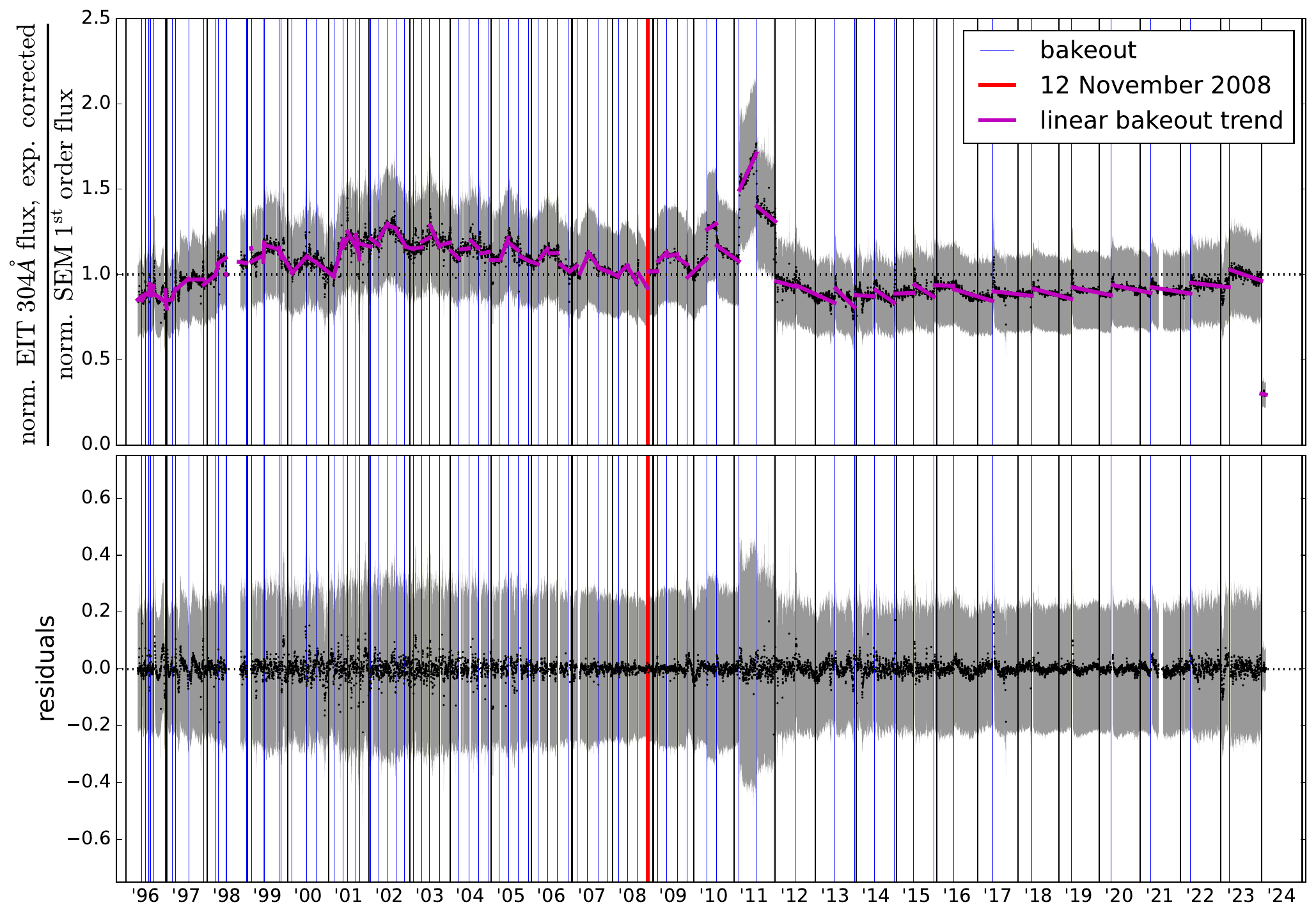}
\caption{Upper panel: The ratio of the normalized EIT $304$\AA\ light curve to the CELIAS/SEM $1^{\mathrm{st}}$-order light curve. The linear functions fit to each inter-bakeout period are plotted in pink. The vertical red line marks 12 November 2008, the (rough) time after which the inter-bakeout trends start to get noticeably worse. Note that the SEM time series ends on 8 Feb 2024 (earlier than our EIT light curves), so this ratio plot also ends on 8 February 2024. Lower panel: The residuals of the linear fits.}
\label{fig:linearBakeout}
\end{center}
\end{figure*}

We propagate the uncertainties in the parameters of the best-fit exponential functions through to the uncertainties on the corrected EIT $304$\AA\ light curve data points analytically:
\begin{equation}
    F_{\mathrm{exp.\ corrected}} = \frac{F}{\exp{[A(t-t_{\mathrm{ref}}) + B]}}
\end{equation}

\begin{multline*}
    \sigma_{F\ \mathrm{exp.\ corrected}}^2 = \left( \frac{\partial F_{\mathrm{exp.\ corrected}}}{\partial F}\right)^2 \sigma_F^2 + \left( \frac{\partial F_{\mathrm{exp.\ corrected}}}{\partial A}\right)^2 \sigma_A^2 + \\
    \left( \frac{\partial F_{\mathrm{exp.\ corrected}}}{\partial B}\right)^2 \sigma_B^2,
\end{multline*}
    
where $A$, $B$, $\sigma_A$, and $\sigma_B$ are given for both exponential fits in Table~\ref{tab:expParams}.




After correcting for the exponential trend, we must next correct for the inter-bakeout trends, which affect all four EIT bands. We make the assumption that each inter-bakeout trend can be modeled by a linear offset to the EIT light curve. There are 84 bakeouts that occur during our light curve baseline, plus one currently unexplained flux discontinuity between 29 and 30 December 2023; these mark the boundaries between each inter-bakeout segment. Using linear least squares fitting, we fit a line to each inter-bakeout segment of the EIT $304$\AA\ to CELIAS/SEM light curve ratio. We then divide this line out of all four of the EIT light curves. 

We once again propagate the uncertainties on the best-fit line through to the uncertainties on the corrected light curves analytically:

\begin{equation}
    F_{\mathrm{bakeout\ corrected}} = \frac{F}{m(t-t_{\mathrm{ref}}) + b},
\end{equation}

where $F$ is given by the $F_{\mathrm{exp.\ corrected}}$ in the case of the 304\AA\ light curve, and by the uncorrected $F$ for the other three bands, and $t_{\mathrm{ref}}$ is the start time of each inter-bakeout segment. Because there are 86 segments, we do not enumerate all the best-fit $m$ and $b$ values here. Meanwhile, for each inter-bakeout segment, the corrected flux uncertainty is given by

\begin{multline*}
    \sigma_{F\ \mathrm{bakeout\ corrected}}^2 = \left( \frac{\partial F_{\mathrm{bakeout\ corrected}}}{\partial F}\right)^2 \sigma_F^2 + \left( \frac{\partial F_{\mathrm{bakeout\ corrected}}}{\partial m}\right)^2 \sigma_m^2 +\\
    \left( \frac{\partial F_{\mathrm{bakeout\ corrected}}}{\partial b}\right)^2 \sigma_b^2.
\end{multline*}


(Implicit in this procedure is the assumption that there is a linear relationship between all four of the EIT light curves; see below for further discussion.) Figure~\ref{fig:linearBakeout} shows the linear fits to each inter-bakeout segment of this ratio and the residuals in this ratio after the linear trends are removed.

We note here that the inter-bakeout trends only cause the EIT $304$\AA\ to CELIAS/SEM light curve ratio to diverge significantly from 1 from roughly 2008 onward. We therefore produce two versions of our final light curves: in version (1), the inter-bakeout trend correction has been done for the entire mission lifetime, and in version (2), it has only been done from 12 November 2008 onward. Because the inter-bakeout trend correction increases the uncertainty on the corrected light curve data points, the version (1) has approximately homoscedastic uncertainties. Version (2) has very diferent uncertainty characteristics pre- and post-2008, but the uncertainties before 2008 are much smaller. Either version could be more appropriate for various types of Sun-as-a-star analysis, depending on the desired uncertainty properties.

\subsubsection{Additional vertical offset correction for 284\AA, 195\AA, and 171\AA\ bands}

This procedure is sufficient to correct the inter-bakeout trends in the EIT 304\AA\ band light curve, as the residuals in Figure~\ref{fig:linearBakeout} show. However, even after dividing out the linear trends, the other three light curves are still noticeably discontinuous across the post-November 2008 bakeouts, particularly in the period between 2008 and 2012. To address these discontinuities, we adjust our systematics model to allow a constant vertical offset to each other EIT light curve (284\AA, 195\AA, 171\AA) for each inter-bakeout period. 

To find the appropriate vertical offset, we start by reiterating our hypothesis that the $284$\AA, $195$\AA, and $171$\AA\ light curves have a linear relationship to each other, at least to first order. Figure~\ref{fig:bakeoutLineup} shows this linear relationship: here, we plot each of the version (2) light curves against each other (that is, these light curves have had the linear inter-bakeout trends based on the the $304$\AA\ to CELIAS/SEM ratio removed, but only from 12 November 2008 onward). For clarity, we do not plot the uncertainties of the data points. 

The line in each subplot of Figure~\ref{fig:bakeoutLineup} represents the best straight-line fit to the pre-12 November 2008 data. We note that all three light curves have been normalized by their median value, to make it easier to set sensible priors on the fit parameters. Because our light curve fluxes are uncertain in all three bands, we specify a model where each two-dimensional projection of this best-fit line has three parameters: a slope $m$, an intercept $b$, and an intrinsic scatter orthogonal to the line, $V$. We perform the fit to each two-dimensional projection with the MCMC routine \texttt{emcee} in a transformed parameter space $\theta, b\cos{\theta}, V$, where $\theta = \arctan{m}$, following the recipe given in \cite{hogg2010}, sections 7 and 8 (specifically, the likelihood function for the fit is given by their equation 35). The priors for this fit are given in Table~\ref{tab:MCMCpriors}, and the parameters of the best-fit line are given in Table~\ref{tab:bestLine}.

\begin{table}
\caption{Priors for the MCMC fit of the linear model with scatter to the relationship between the pre-12 November 2008 EIT $171$\AA, $195$\AA, and $284$\AA\ normalized light curves.} 
\centering 
\begin{tabular}{l l}
\hline\hline 
Fit parameter & Prior \\ [0.5ex] 
\hline 
  $\theta$ &  $U\sim[0, \pi/2]$\\
  $b\cos{\theta}$ & $U\sim[-10,10]$ \\
  $V$ & $U\sim[0, 5]$\\
  [1ex]
\hline\hline 
\end{tabular}
\label{tab:MCMCpriors} 
\end{table}

\begin{table*}
\caption{Parameters of the best-fit line to the relationship between the pre-12 November 2008 EIT $171$\AA, $195$\AA, and $284$\AA\ normalized light curves.} 
\centering 
\begin{tabular}{l l l l}
\hline\hline 
Projected variables & Slope $m$ & Intercept $b$ & Scatter $V$ \\ [0.5ex] 
\hline 
  $x = 284$\AA\, $y=195$\AA\ & $0.707\pm0.002$ & $0.368\pm0.002$ & $0.00161 \pm 0.00006$\\
  $x = 284$\AA\, $z=171$\AA\ & $0.349\pm0.002$ & $0.699\pm0.002$ & $0.00205 \pm 0.00007$\\
  $y = 195$\AA\, $z=171$\AA\ & $0.496\pm0.002$ & $0.516\pm0.002$ & $0.00112 \pm 0.00004$\\
  [1ex]
\hline\hline 
\end{tabular}
\label{tab:bestLine} 
\end{table*}

We assume that, if the bakeouts are appropriately corrected, that all of the data should lie as close to the best-fit pre-12 November 2008 line as possible. For a given inter-bakeout segment, we calculate the shortest three-dimensional vector between the point $(m_{171}, m_{195}, m_{284})$, where $m_{171}$ is the median value of the $171$\AA\ light curve during segment, and the three-dimensional line of best fit to the pre-12 November 2008 data, the two-dimensional projections of which are plotted in Figure~\ref{fig:bakeoutLineup}. The three components of this vector are the vertical offsets we apply to each light curve to in order to minimize discontinuities across bakeouts. 

The left panel of Figure~\ref{fig:bakeoutLineup} shows this procedure graphically for the inter-bakeout segment beginning 22 July 2011 and ending 6 January 2012. The blue symbols show the light curve data from this segment before the vertical offsets are calculated and applied; the magenta symbols show the data after the correction. The right panel of Figure~\ref{fig:bakeoutLineup} shows the EIT 171\AA\ light curve during early Solar cycle 24, when the bakeout discontinuities are most extreme, before and after the vertical offsets are applied. Applying the offsets significantly ameliorates the discontinuities.


For the version (1) light curves, we repeat this vertical offset correction procedure for every inter-bakeout segment where it is possible. It is not possible for the inter-bakeout segments in 1996 and 1997 where there is no 284\AA\ data due to the filter pinhole problem, because the three-dimensional point $(m_{171}, m_{195}, m_{284})$ is not defined for these segments. For the version (2) light curves, we repeat the vertical offset correction procedure for only the inter-bakeout segments after 12 November 2008. 

\begin{figure}
\begin{center}
\includegraphics[width=0.48\textwidth]{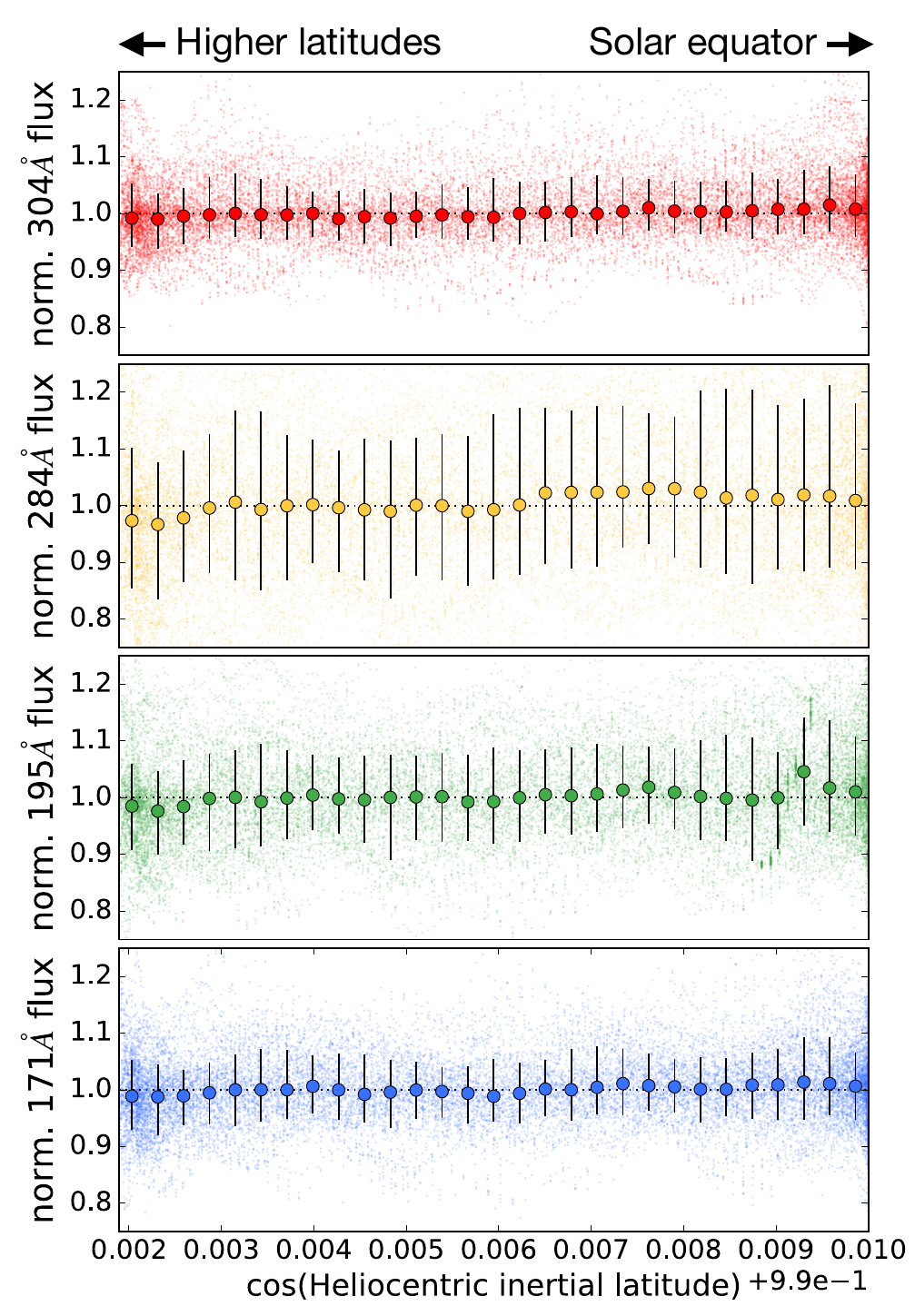}
\caption{Small points: The four EIT version (2) light curves, plotted against the cosine of SOHO's heliocentric inertial latitude. $\cos{\mathrm{(lat)}} = 1$, at the right-hand side of the plot, corresponds to a SOHO heliocentric latitude of $0^{\circ}$, or a viewing angle directly onto the equator. To prevent this plot from being dominated by the solar activity or rotation cycles, we divide each light curve into segments according to SOHO's heliocentric latitude and normalize the flux per segment (see text for details).}
\label{fig:heliocentricLat}
\end{center}
\end{figure}

\subsubsection{Viewing angle correction}\label{subsubsec:viewingAngle}

Lastly, we introduce another small correction to the bakeout-corrected light curves, to account for SOHO's variable viewing angle on the Sun throughout its orbit. Because of the tilt of the ecliptic plane with respect to the solar equator, and the tilt of SOHO's orbit with respect to the ecliptic plane, SOHO's heliocentric latitude varies between approximately $-7.5^{\circ}$ and $+7.5^{\circ}$. Naively, we would expect that, because the solar corona is brighter at the Sun's active latitudes than at the poles, the EIT light curve points taken at higher latitudes, where more of the pole is visible, should be fainter.

Figure~\ref{fig:heliocentricLat} shows the four EIT light curves, sorted by heliocentric inertial latitude. In order to prevent the solar activity cycle or the solar rotation from dominating this plot, we divide up the EIT light curves into segments corresponding to one quarter of SOHO's orbit, and normalize each segment by its median flux value. In the first quarter, SOHO sweeps from heliocentric latitude $0^{\circ}$ to $+7.5^{\circ}$; in the second quarter, from $+7.5^{\circ}$ back to $0^{\circ}$, and so on.

To establish whether there is a trend in the data with heliocentric latitude, we fit the light curves as plotted in Figure~\ref{fig:heliocentricLat} with three competing models and evaluate the Bayesian evidence for each. The first model has no dependence on heliocentric latitude,

\begin{equation}
    F  = k.
\end{equation}

The second is linear in the absolute value of heliocentric latitude $l$,

\begin{equation}
    F = m_l|l| + b_l.
\end{equation}

The third is linear with respect to the cosine of heliocentric latitude,

\begin{equation}
    F = m_c(\cos{l} - \cos{l_0}) + b_c,
\end{equation}

where $l_0$ is the minimum heliocentric latitude across the time series.

We use the dynamic nested sampler from the \texttt{python} package \texttt{dynesty} (\citealt{speagle2020}) to perform the fits and the evidence calculation. The priors for these fits are given in Table~\ref{tab:dynestypriors} (although the outcome of the model selection is not sensitive to the choice of boundaries on these priors). We repeat this procedure for both version (1) and version (2) of the light curves. 

\begin{table}
\caption{Priors for the nested sampling fit of each normalized light curve as a function of heliocentric latitude. The outcome of model selection is not sensitive to the boundaries on these priors.} 
\centering 
\begin{tabular}{l l}
\hline\hline 
Fit parameter & Prior \\ [0.5ex] 
\hline 
  constant model & \\
  $k$ &  $U\sim[0.95, 1.05]$\\
  \hline
  linear model &\\
  $m_l$ & $U\sim[-0.025, 0.025]$ \\
  $b_l$ & $U\sim[0.95, 1.05]$\\
  \hline
  linear in $\cos{l}$ &\\
  $m_c$ & $U\sim[-10,10]$ \\
  $b_c$ & $U\sim[0.95, 1.05]$\\
  [1ex]
\hline\hline 
\end{tabular}
\label{tab:dynestypriors} 
\end{table}

The natural logarithm of the Bayesian evidence ratios for the various models are presented in Table~\ref{tab:Zfac}. In almost every case, the model that is linear in the cosine of heliocentric latitude is decisively preferred over the other two. The exception is the version (2) 304\AA\ light curve, for which the model that is linear in heliocentric latitude is preferred. In every case, the Bayes factor, equal to $Z_{\mathrm{linear\ cos}}/Z_{\mathrm{linear}}$, has a magnitude greater than 10, meaning that the preference is strong. 

Finally, we divide out the preferred model from each light curve, again propagating the uncertainties in the best-fit model parameters analytically to the corrected light curve data points. For all but the version (2) 304\AA\ light curves, the corrected flux is then given by

\begin{equation}
    F_{\mathrm{viewing\ angle\ corrected}} = \frac{F_{\mathrm{bakeout\ corrected}}}{m_c (\cos{l}-\cos{l_0}) + b_c}.
\end{equation}

For the version (2) 304\AA\ light curves, it is given by

\begin{equation}
    F_{\mathrm{viewing\ angle\ corrected}} = \frac{F_{\mathrm{bakeout\ corrected}}}{m_l |l| + b_l}.
\end{equation}

The corrected uncertainty, finally, is given by 

\begin{multline*}
    \sigma_{F\ \mathrm{viewing\ angle\ corrected}}^2 = \left( \frac{\partial F_{\mathrm{viewing\ angle\ corrected}}}{\partial F_{\mathrm{bakeout\ corrected}}}\right)^2 \sigma_{F\ \mathrm{bakeout\ corrected}}^2 +\\
    \left( \frac{\partial F_{\mathrm{viewing\ angle\ corrected}}}{\partial m}\right)^2 \sigma_m^2 + \left( \frac{\partial F_{\mathrm{viewing\ angle\ corrected}}}{\partial b}\right)^2 \sigma_b^2,
\end{multline*}

where $m, b$ = $m_l, b_l$ for the version (2) 304\AA\ cases and $m_c, b_c$ for all other cases.



\begin{table*}
\caption{A comparison of the Bayesian evidences for the three models: ``constant'', which is independent of heliocentric latitude; ``linear'', which is linear in heliocentric latitude; and ``linear cos'', which is linear in the cosine of heliocentric latitude.  We present the logarithm of the evidence ratios of the ``linear cos'' model to the other two; when this number is positive, as it is in most cases here, it means that the linear cos model is preferred. In every case, the linear cos model is preferred to the constant model. In every case except the version (2) 304\AA\ light curve, the linear cos model is also preferred to the linear model. The uncertainties are the statistical uncertainties reported by \texttt{dynesty}.} 
\centering 
\begin{tabular}{l l l}
 wavelength& $\ln{[Z_{\mathrm{linear\ cos}}/Z_{\mathrm{constant}}
]}$ &  $\ln{[Z_{\mathrm{linear\ cos}}/Z_{\mathrm{linear}}]}$\\ [0.5ex] %
\hline
Version 1 & (bakeouts corrected over whole mission lifetime)& \\
\hline
  171\AA\ & $19.01 \pm 0.07$ & $4.06 \pm 0.09$\\
  195\AA\ & $73.81 \pm 0.07$ & $6.53 \pm 0.09$\\
  284\AA\ & $181.40 \pm 0.07$ & $48.02 \pm 0.09$\\
  304\AA\ & $80.92 \pm 0.08$ & $6.69 \pm 0.09$\\
  [1ex]
\hline
Version 2 & (bakeouts corrected from 12 November 2008 onward)& \\
\hline
  171\AA\ & $35388580 \pm 40$ & $2779730 \pm 30$\\
  195\AA\ & $74364900 \pm 60$ & $4646920 \pm 70$\\
  284\AA\ & $15234477 \pm 9$ & $1759480 \pm 10$\\
  304\AA\ & $22603850 \pm 20$ & \textbf{$-366600 \pm 20$}\\
  [1ex]
\hline
\end{tabular}
\label{tab:Zfac} 
\end{table*}

\section{Results}
\label{sec:results}
\begin{figure*}
\begin{center}
\includegraphics[width=\textwidth]{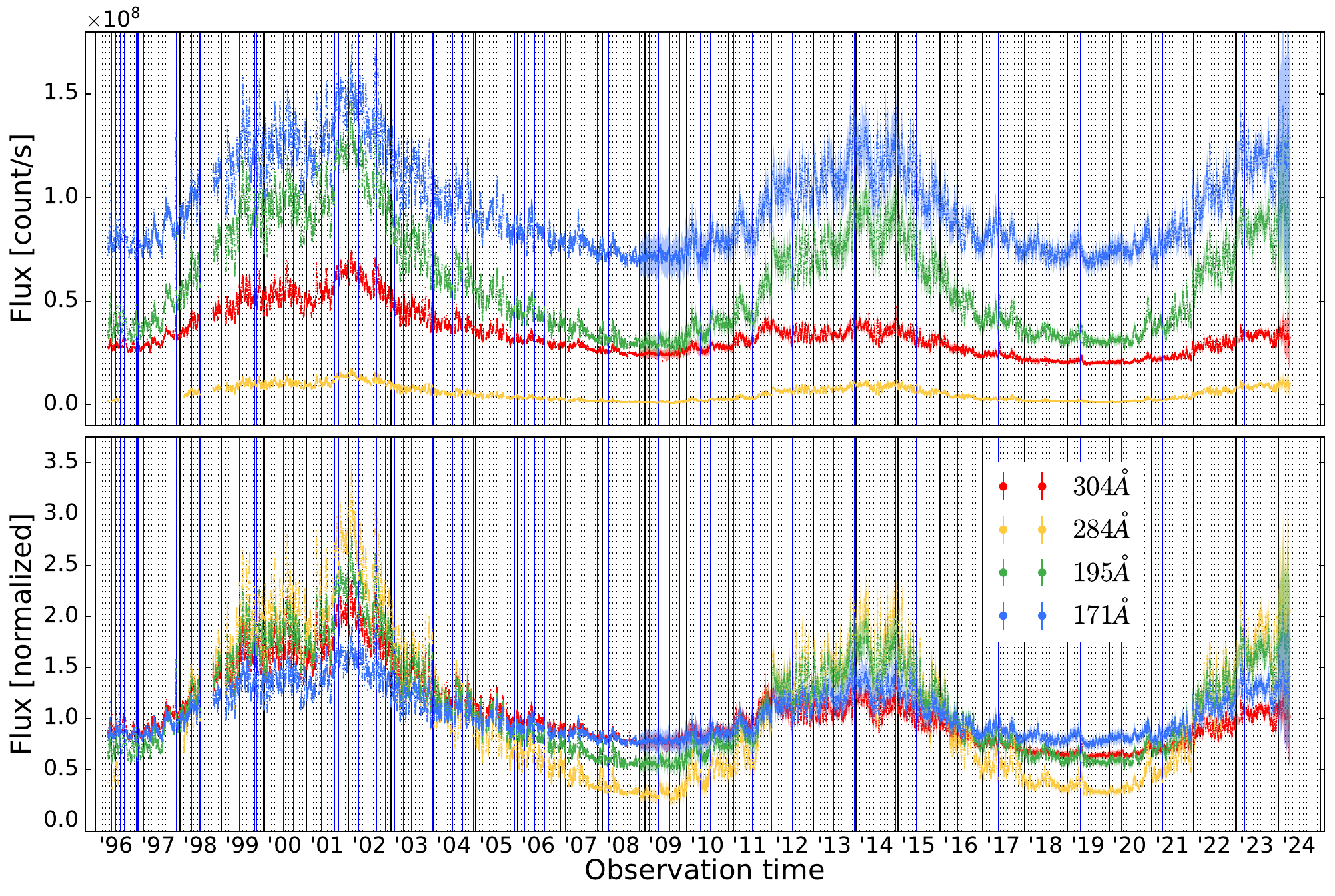}
\caption{The final EIT light curves, corrected for instrumental systematics and for EIT's viewing angle. These are the version (2) light curves, for which the linear bakeout trend correction is only performed after 12 November 2008; as a result, the light curve uncertainties after this date are much larger than before this date. Upper panel: The un-normalized light curves in their original units. Lower panel: The same light curves, each normalized by its median value.}
\label{fig:LCs}
\end{center}
\end{figure*}

Thus, we produce our final EIT light curves, in all four bands. We publish two versions of each Sun-as-a-star light curve: version (1), uniformly corrected for instrumental systematics across the whole mission lifetime (except the inter-bakeout segments in 1996 and 1997 with no 284\AA\ data, where vertical inter-bakeout segment offsets cannot be calculated, as noted above), and version (2), corrected for bakeout systematics only after 12 November 2008. For completeness, we also publish versions (1) and (2) without the final correction for viewing angle outlined in Section~\ref{subsubsec:viewingAngle}. Table~\ref{tab:versions} summarizes the differences between the various versions and suggests use cases for each.

\begin{table*}
\caption{Description of the four published final versions of each EIT light curve.} 
\centering 
\begin{tabular}{l l l}
\hline\hline 
Version & Filename string & Description \& suggested use cases \\ [0.5ex] 
\hline 
  \makecell{version (1), viewing angle-corrected\\ \hspace{1.65cm} no viewing angle correction} & \makecell{corrected\_all\_heliocentriccorr \\ corrected\_all} & \makecell{Uniformly corrected for instrumental \\ systematics across whole mission \\lifetime; $\sim$ homoscedastic uncertainties. \\ Suitable for e.g. testing pipelines that \\ will be applied to long-baseline stellar \\ light curve data with homoscedastic\\ uncertainties. \\Version without viewing angle correction\\ could be used to study systematic trends\\ in data from other SOHO instruments.}  \\
  \makecell{version (2), viewing angle-corrected \\ \hspace{1.65cm} no viewing angle correction} & \makecell{corrected\_2008onwards\_heliocentriccorr \\ corrected\_2008onwards} & \makecell{Corrected for bakeouts from 12 Nov \\2008 onward only; heteroscedastic \\ uncertainties, where pre-12 Nov 2008 \\uncertainties are much smaller. \\ Suitable for e.g. detailed analysis of how \\ solar activity features appear in light \\ curves, and for analysis of how light curve\\ precision affects our ability to infer or\\ reconstruct such features. \\  Version without viewing angle correction\\ could be used for more detailed study of\\ EUV flux incident on Earth.} \\
\hline\hline 
\end{tabular}
\label{tab:versions} 
\end{table*}

Figure~\ref{fig:LCs} shows the final version (2), viewing angle-corrected light curves in all four EIT bands. The upper panel shows the un-normalized light curves in their original units of [count/s], so the different efficiencies of the four EIT bandpasses are clearly visible (see Figure~\ref{fig:EITbandpasses})---the 171\AA\ bandpass has the highest transmission, and the 284\AA\ light curve has the lowest. The lower panel shows the same light curves normalized by their respective median values.

The overall modulation due to solar cycles 23 (mid-1996 - late 2008), 24 (late 2008 - late 2019), and 25 (late 2019 onward) is clearly visible in all four light curves, with the maximum in cycle 23 being stronger than the maximum in cycle 24, as expected from other solar observations. Over the solar cycle, the 284\AA\ band light curve has the highest peak-to-peak fractional variability (320\%), followed by the 195\AA\ band (230\%), the 304\AA band (160\%), and finally the 171\AA band (130\%). The solar rotation cycle is also visible in all four light curves, with the strongest fractional variability again apparent in the 284\AA\ band, which traces active regions most directly.

It is clear from Figure~\ref{fig:LCs} that residual bakeout discontinuities persist in the version (2) light curves; for example, the bakeout ending 1 September 2001 coincides with a discontinuity in all four light curves. This discontinuity is significantly ameliorated in the version (1) light curves due to their larger uncertainties in the first half of the mission lifetime.

Figure~\ref{fig:LCs} also suggests, tentatively, that there may be an uncorrected systematic decline in the 304\AA\ light curve from at least 2008 onward. This decline can be seen by comparing the normalized 304\AA\ light curve to the normalized 171\AA\ light curve. Over solar cycle 23, the 304\AA\ flux is generally greater than the 171\AA\ flux, but the two light curves level out to be roughly the same at solar minimum in 2008-2009, and after $\sim2011$, the 304\AA\ flux is generally lower than the 171\AA\ flux. 

Because we calibrate the EIT 304\AA\ light curve under the assumption that the degradation-corrected CELIAS/SEM $1^{\mathrm{st}}$-order light curve represents the ground truth, this yet-uncorrected degradation may indicate that our linear degradation correction (see Figure~\ref{fig:SEMcorrection}) is not sufficient over the latter half of the mission. We therefore encourage caution when using our 304\AA\ EIT light curve, particularly over long baselines.

\begin{figure*}
\begin{center}
\includegraphics[width=\textwidth]{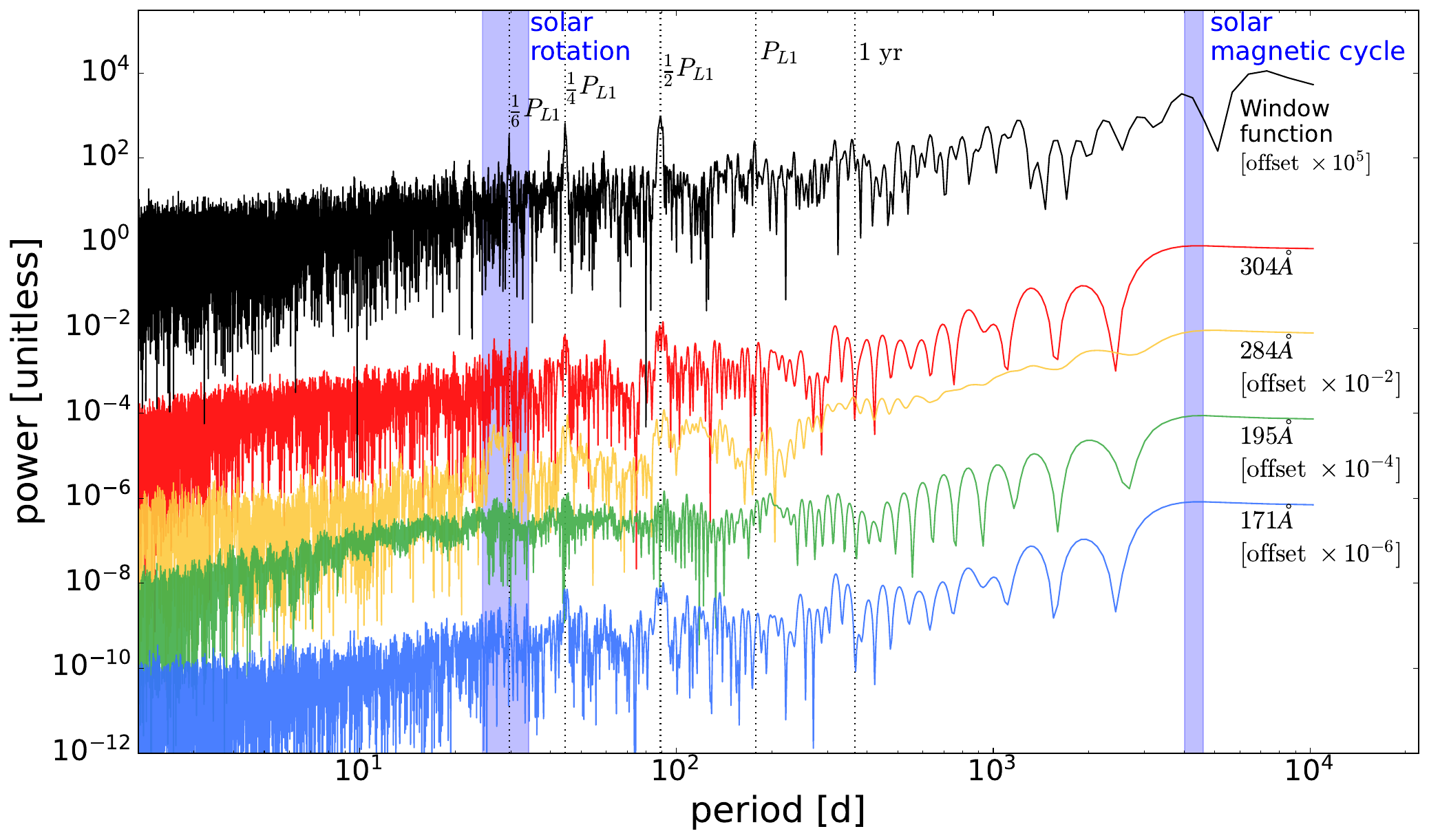}
\caption{Lomb-Scargle periodograms of the version (2) light curves presented in Figure~\ref{fig:LCs}, as well as of the window function of the observations. A multiplicative offset has been applied to each periodogram to render them all visible on the same plot. The window function has significant power at frequencies corresponding to $1/2$ and $1/4$ of SOHO's orbital period, because EIT does not take data during SOHO's keyhole periods, which occur twice per orbit. There is also excess power in the window function at $1/6$ of SOHO's orbital period. All four light curves have excess power at the solar rotation period, while the window function does not.}
\label{fig:periodogram}
\end{center}
\end{figure*}

Figure~\ref{fig:periodogram} shows Lomb-Scargle periodograms \citep{lomb1976,scargle1982}, computed with \texttt{astropy.timeseries.LombScargle}, of the four EIT light curves and of the window function imposed by EIT's observation schedule (i.e., a constant time series with the same ``observation times'' as the EIT light curve). The window function has excess power at frequencies corresponding to 1 year (Earth's orbital period, and therefore L1's orbital period) and harmonics of SOHO's orbital period around L1 ($\mathrm{P}_{\mathrm{SOHO}}\sim 178$ days). The EIT light curves also have excess power at these frequencies, particularly $\mathrm{P}_{\mathrm{SOHO}}/2$ and $\mathrm{P}_{\mathrm{SOHO}}/4$, which makes sense given the symmetry of the orbit and the twice-per-orbit keyhole periods. The EIT light curves also have excess power at and around the solar rotation period of $\sim27$ days. 

We next evaluate whether the solar rotation period could be recovered from our light curves under observational constraints similar to those of a short-baseline photometric monitoring campaign, such as an exoplanet transit survey. We split each of our light curves up into 1-year segments by calendar year of observation, yielding 29 segments per light curve. (Note that the first and last segments are shorter than one year--the 1996 segment only includes data from April to December, and the 2024 segment only includes data from January to April.) Next, following the method of \cite{angus2018} for inferring probabilistic stellar rotation periods from light curve data and using the python package \texttt{celerite} \citep{dfm2017}, we fit each segment using a Gaussian process model with a quasi-periodic covariance kernel, the form of which is given by \cite{dfm2017} equation 56:

\begin{equation}
    \kappa(\tau) = \frac{B}{2+C}e^{-\tau/L}\left[ \cos{\frac{2\pi\tau}{P_{\mathrm{rot}}}} + (1+C) \right]
\end{equation}

Here, $P_{\mathrm{rot}}$ is the rotation period; the other parameters have less direct physical interpretation, but (depending on their numerical values) the covariance function describes a time series where points separated by $nP_{\mathrm{rot}}$ are correlated, with the strength of the correlation decaying for increasing $n$.

\begin{table}
\caption{Priors for the MCMC fit of the Gaussian process model to each light curve segment. The upper bound of the rotation period prior corresponds to SOHO's orbital period around L1.} 
\centering 
\begin{tabular}{l l}
\hline\hline 
Fit parameter & Prior \\ [0.5ex] 
\hline 
  $\ln{B}$ &  $U\sim[-10, 10]$\\
  $\ln{C}$ & $U\sim[-10,10]$ \\
  $\ln{L}$ & $U\sim[-10,10]$\\
  $\ln{P_{\mathrm{rot} [d]}}$ & $U\sim[0, \ln{178}]$\\
  [1ex]
\hline\hline 
\end{tabular}
\label{tab:period_priors} 
\end{table}

For each year-long EIT light curve segment, we fit the parameters of this covariance function using \texttt{emcee} \citep{dfm2013}. The priors of the \texttt{emcee} fit are given in Table~\ref{tab:period_priors}. We consider a fit to be ``converged'' if the maximum-a-posteriori solution isn't within 1 day of either the lower or upper limit of the $\ln{P_{\mathrm{rot}}}$ prior. 

\begin{figure*}
\begin{center}
\includegraphics[width=\textwidth]{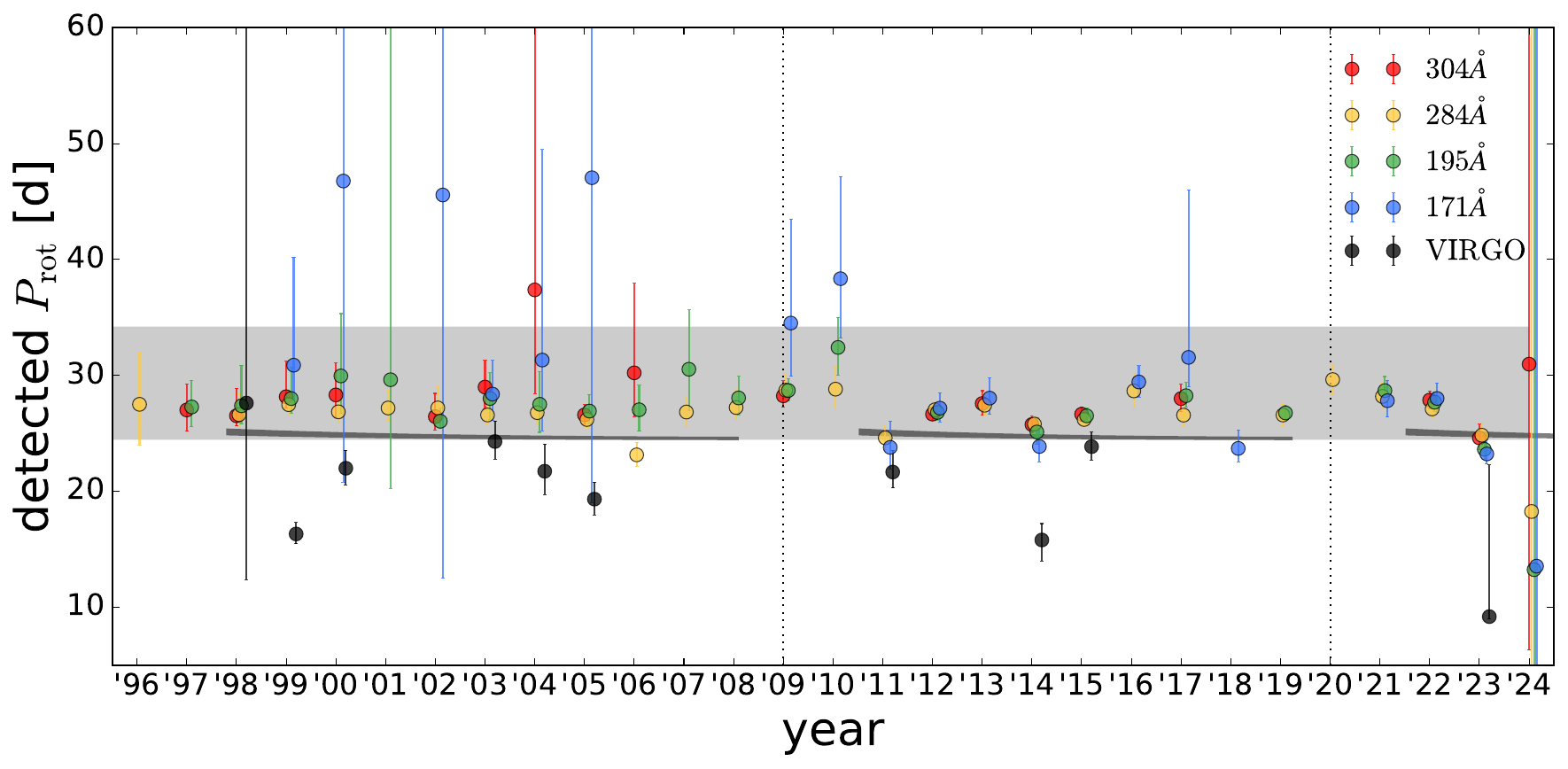}
\caption{The recovered rotation periods from a Gaussian process fit to segments of each EIT version (2) light curve and the VIRGO light curve, split up by calendar year of observation. Each band has a small x-offset for plotting clarity. The gray shaded band shows the range of the true solar rotation period, from equator to pole, calculated analytically from the solar differential rotation equation given by \cite{snodgrass1990}. The near-horizontal black curves represent the average behavior of migrating sunspot groups over the solar activity cycle, as calculated by \cite{jiang2011}. The true solar rotation period can be recovered from the 284\AA\ light curve in 26 out of 29 years, and from the VIRGO light curve in only 3 out of 29 years.}
\label{fig:periods}
\end{center}
\end{figure*}

We repeat this experiment for the VIRGO total solar irradiance time series as well. The results for EIT and VIRGO are plotted in Figure~\ref{fig:periods}. We only show data points for which the Gaussian process model fit converged: 18/29 segments for the 304\AA\ light curve, 27/28 segments for 284\AA\ (recall that there is no usable data in the 284\AA\ band in 1997 due to pinholes), 23/29 segments for 195 \AA, 19/29 segments for 171\AA, and 10/29 segments for VIRGO. The shaded horizontal band in Figure~\ref{fig:periods} shows the range of the true solar rotation period, from equator to pole, calculated analytically from the solar differential rotation equation given by \cite{snodgrass1990}. The near-horizontal black curves represent the average rotation period behavior of sunspot groups over the solar activity cycle, as calculated by \cite{jiang2011}. 

For the 284\AA\ light curve, the maximum a-posteriori rotation period value is $1\sigma$ consistent with the true solar rotation for 26 out of 28 segments. The only years for which the true solar rotation period is not recoverable from the 284\AA\ light curve are 2006, when the maximum a-posteriori rotation period slightly underestimates the true value, and 2018, when the maximum a-posteriori rotation period abuts the upper limit of the rotation period prior. The uncertainty on the best-fit period in 2024 is extremely large, because the 2024 baseline is only $\sim3$ months.

Meanwhile, the true solar rotation is recovered for 23/29 segments of the 195\AA\ light curve, 18/29 segments of the 171\AA\ light curve, 18/29 segments of the 304\AA\ light curve, and only 3/29 segments of the VIRGO light curve. In most cases where the VIRGO fits converge, they underestimate the true solar rotation period. There is nothing obviously wrong with the fits that converge to a low value, and changing the bounds on the parameter priors does not significantly affect the results.

The difference between the success of this exercise in the EUV vs. the optical is stark: for the 284\AA\ band, which directly traces coronal activity, the solar rotation period is recoverable 93\% of the time, including in 1996, where there are only $\sim8.5$ months of light curve data. Meanwhile, for the VIRGO TSI time series, which is dominated by optical light, the solar rotation period is only recoverable 10\% of the time. Changing the prior ranges for $\ln{B}$, $\ln{C}$, and $\ln{L}$ has no significant effect on this result.

\section{Discussion \& Conclusions}
\label{sec:discussion}


Here, we present Sun-as-a-star light curves from the SOHO EIT images in four narrow extreme ultraviolet bands, tracing solar activity in the chromosphere (at 304\AA) and corona (at  284\AA, 195\AA, and 171\AA) across 28 years and 2.5 solar activity cycles. These light curves are available for download at \href{https://doi.org/10.5281/zenodo.15474179}{https://doi.org/10.5281/zenodo.15474179}. For the first time, it is possible to directly compare EIT's extremely detailed picture of activity on the Sun to observations of other Sun-like stars. 

We find, furthermore, that these EUV light curves are better tracers of the solar activity cycle than the Sun's optical light curve. The Sun's fractional variability over the solar cycle is much higher at UV and X-ray wavelengths than at optical wavelengths. Our EIT 284\AA\ light curve has a peak-to-peak fractional variability of 320\% over SOHO's baseline; over the same baseline, the fractional variability in the VIRGO total solar irradiance time series is only 0.3\%. 

The EIT light curves are also substantially more sensitive to the rotation of solar activity features than optical light curves. To test this, we divided our light curves up into year-long segments and attempted to infer the solar rotation period from each. We find that we can accurately recover the solar rotation period for 26/28 year-long segments of the EIT 284\AA\ light curve (93\%), compared to only 3/29 year-long segments of the VIRGO light curve (10\%). This result agrees nicely with recent other work on the detectability of solar rotation at different wavelengths; \cite{vasilyev2025}, for example, use the SATIRE-S solar spectral irradiance data to calculate the probability of accurately recovering the solar rotation period from light curve data at wavelengths between 2000\AA\ and 10000\AA, concluding that at optical wavelengths ($\geq 400$nm), the probability of detecting the true $P_{\mathrm{rot}}$ at a random phase in the solar cycle is less than 20\%, while at UV wavelengths ($< 400$nm), the probability is as high as 80\%.

These results align with expectations from the literature on stellar activity: a star's UV emission is known to trace its activity state and phase in its magnetic cycle (if it has one). The activity state, in turn, is important to know, because it can affect the precision and accuracy of radial velocity measurements of the star (see e.g. \citealt{santos2010, lovis2011, dumusque2011, meunier2024}) and because it determines the strength of the stellar wind (see e.g. \citealt{vidotto2012}). In other words, the activity state, which is directly traced by the UV emission, bears directly on both the detectability and habitability of any planets in the stellar system. The EUV flux itself, and its variability, are also important to understand, as they are likely to have significant effects on the chemistry and thermal structure of the atmospheres of any surrounding planets; see \cite{climatereport2012} for a detailed interdisciplinary review of the effects of solar variability on Earth's climate.

However, ultraviolet observations of stars are much scarcer than optical observations, especially in the current age of high-quality, long-baseline, high-cadence photometric observations of hundreds of thousands of stars taken by exoplanet transit surveys, including Kepler and TESS. For example, broad-band near-UV photometry (1350-2750\AA) from the Galaxy Evolution Explorer (GALEX; \citealt{martin2005}) mission's All-Sky Imaging Survey has been used to study the UV emission of nearby M-dwarfs \citep{shkolnik2011, stelzer2013} and Sun-like stars \citep{findeisen2011} and to compare this UV emission to optical activity indicators from spectra, including H$\alpha$ emission and $R'_{\mathrm{HK}}$, finding general positive correlation but large scatter. \cite{findeisen2011} conclude that, for their sample of Sun-like stars, roughly half of this scatter is attributable to observational uncertainty and half to long-term, activity-related UV variability across their stellar sample. At shorter wavelengths, EUV and X-ray photometric measurements have also been taken for hundreds of Sun-like stars in order to study the relationship between high-energy emission and stellar age (e.g. \citealt{ribas2005}) and the more general age-activity-rotation relationship (see e.g. \citealt{pizzolato2003}), but little has been directly measured about the high-energy variability of these stars over timescales comparable to the solar activity cycle. Compared to the $\mathcal{O}(10^5)$ stars observed through photometric time series, this represents a very small and specific subset.

Spectral observations of active stars in the UV are even rarer, but also indicate that optical photometry cannot paint a full picture of stellar activity. \cite{france2016} measure the X-ray to mid-IR spectral energy distributions of 11 nearby K and M-type exoplanet host stars and observe that many of the stars in their sample have strong UV flare emission without any accompanying signs of strong stellar activity at optical wavelengths. \cite{youngblood2017} use the optical and UV spectra from this stellar sample to derive scaling relations between UV emission, energetic particle flux from stellar flares, and proxy optical activity indicators visible from the ground. These relations allow estimation of the variability of UV emission based on the variability of optical indicators, but the expense of taking UV spectra means that, once again, UV variability over the stellar activity cycle has been beyond direct observational reach for statistical samples of stars thus far. The upcoming Mauve mission \citep{mauve} will take near-UV spectra (2000-7000\AA) of hundreds of stars across a 3-year mission baseline, enabling studies of ultraviolet variability on timescales of years.

There have also been significant efforts to predict UV emission from optical light curves. \cite{cegla2014}, for example, used GALEX observations of the Kepler field to study the relationship between a star's UV emission and the variability in its optical light curve, in an effort to identify which stars are magnetically quiet before investing in spectroscopic observations. They find that, for magnetically active stars (those above the ``flicker floor'', \citealt{bastien2013}), there is significant scatter in the relationship they derive between photometric variability and the UV emission-predicted spectroscopic activity indicators, which they attribute to ``variations in the phase and strength of the stellar magnetic cycle" of stars in their sample. 


We present the EIT Sun-as-a-star light curves to aid these efforts to relate optical and UV photometry under the muddling influence of stellar activity cycles and learn more about the activity state of the enormous population of stars with long-baseline optical light curves but no detailed spectroscopic monitoring history. The EIT light curves, combined with the VIRGO total solar irradiance time series, paint a detailed picture of the relationship between the Sun's EUV emission and its optical light curve over two and a half solar magnetic cycles; represent a long-baseline supplement to ongoing but relatively short-baseline spectroscopic optical Sun-as-a-star monitoring (see e.g. \citealt{colliercameron2019}); and illuminate ongoing efforts to monitor the activity cycles of other stars at multiple wavelengths (see e.g. \citealt{wargelin2024} for multi-wavelength observations of Proxima Centauri). 

The versions of our EIT light curves without the final viewing angle correction step, meanwhile, may be used in comparison to light curves from other SOHO instruments, or as a long-baseline ($\sim2.5$ activity cycles) complement to detailed spectroscopic observations of the recent EUV radiation flux on Earth, such as those by SORCE/XPS (which operated from 2003-2020) and the Extreme-ultraviolet Variability Experiment \citep{woods2012} aboard the Solar Dynamics Observatory, which launched in 2010.

The library of EIT images from which we derive our light curves, meanwhile, remains an invaluable resource for understanding, in extremely fine detail, the chromospheric and coronal conditions that produce a particular set of optical and EUV light curve data points.

\section*{Acknowledgments}
SOHO is a project of international cooperation between ESA and NASA. The authors thank the reviewer for helpful comments. Emily Sandford received support from the ESA Archival Research Visitor Programme. The authors thank J.B. Gurman for his thorough and helpful comments on the draft, Jean-Philippe Olive and Kevin Schenk for providing additional SOHO and EIT mission information, and Andy To for sharing relevant heliophysics literature. A.M. acknowledges funding from a UKRI Future Leader Fellowship, grant number MR/X033244/1 and a UK Science and Technology Facilities Council (STFC) small grant ST/Y002334/1. L.A.H. is supported through a Royal Society-Research Ireland University Research Fellowship.

This work made use of \texttt{numpy} \citep{numpy}, \texttt{astropy} \citep{astropy1, astropy2, astropy3}, \texttt{SunPy} \citep{sunpy}, \texttt{celerite} \citep{dfm2017},  \texttt{emcee} \citep{dfm2013}, \texttt{scipy} \citep{scipy}, and \texttt{matplotlib} \citep{matplotlib}.

%
\bibliographystyle{aa} 
\bibliography{bib} 
%

\label{lastpage}
\end{document}